\documentclass[]{mn2e} %
\usepackage{graphicx}	
\usepackage{amsmath}	
\usepackage{amssymb}	
\usepackage{bm}   
\def\red{ } 

\def\revone{}

\title[3D Central Molecular Zone]{Three-Dimensional Structure of the Central Molecular Zone}

\author[Y. Sofue]{Yoshiaki Sofue\thanks{E-mail: sofue@ioa.s.u-tokyo.ac.jp} \\ 
Institute of Astronomy, The University of Tokyo, Mitaka, Tokyo 181-0015, Japan }

\date{Accepted; Received YYY; in original form} 
\pubyear{2020}
  
\def\vlsr{V_{\rm LSR}}  \def\vrot{V_{\rm rot}}  \def\Vrot{V_{\rm rot}}  
\def\Msun{M_\odot} 
\def\deg{^\circ} \def\Tb{T_{\rm b}}
\def\Ico{I_{\rm CO}}  \def\Ihi{I_{\rm HI}} 
\def\be{\begin{equation}} \def\ee{\end{equation}}
\def\be{\begin{equation}} \def\ee{\end{equation}}
 \def\XCO{X_{\rm CO}}
\def\Xhi{X_{\rm HI}} \def\Xco{X_{\rm CO}}
\def\kms{km s$^{-1}$}
\def\ekms{{\rm ~km~s^{-1}~}} 
 \def\rmh2{{\rm H_2}} 
\def\Kkms{K \kms }

\def\cos{ {\rm cos}}
\def\be{\begin{equation}} \def\ee{\end{equation}}

\def\thco{$^{13}$CO } \def\twco{$^{12}$CO }
\def\sin{\rm {~sin~}}\def\cos{\rm {~cos~}}
\def\fmol{f_{\rm mol}}
\def\sss{\subsubsection}
\def\dinfty{\infty${\hskip -1.8mm}$\infty}
\def\log{{\rm log}}

\def\Htwo{H$_2$ }
\def\exp{{\rm exp}}
\def\vcut{V_{\rm cut}} 
\def\absv{|V_{\rm lsr}|}
\def\Ico{I_{\rm CO}}
\def\mH{m_{\rm H}}
\def\xcounit{H$_2$ cm$^{-2}$ [K \kms]$^{-1}$ }

\begin{document} 
\maketitle  
\begin{abstract} 
A detailed comparison of HI and CO line cube data of the Galactic Center (GC) region from the archives is obtained.
The central molecular zone (CMZ) is shown to be embedded in the HI disc (central HI zone, CHZ) of radius $\sim 320$ pc and vertical scale height $\sim 70$ pc.
A radio continuum belt is shown to run parallel to molecular Arms I and II. 
The belt draws a double infinity ($\dinfty$) on the sky, connecting Sgr E ($l\sim -1\deg.2)$, C, B1, B2 and Sgr D ($+1\deg.2$), and is interpreted as a warping star-forming ring.
The molecular Arms are closely associated with the HI arms on the longitude-velocity diagram (LVD), showing coherent rigid-body ridges.
Due to the close relationship between HI and CO, the HI line absorption can be used to determine the Arms' position relative to Sgr A, B1, B2 and C.
Combining the trigonometric data of proper motions of Sgr A$^*$ and maser sources of Sgr B2 as well as radial velocities, the 3D velocity vector of Sgr B2 is determined.
From these analyses, the molecular Arm I with Sgr B2 is shown to be located in the near side of Sgr A$^*$, and Arm II with Sgr C in the other side, both composing a pair of symmetrical Arms around the GC.
We present a possible 3D view of Sgr A through E and Arms I and II along with a parameter list.
\end{abstract}
 

\section{INTRODUCTION}

The three-dimensional (3D) structure of the Central Molecular Zone (CMZ) has long been debated, but remains still controversial, because of the edge-on orientation of the central disc of the Galaxy \cite{morris+1996,henshaw+2022}. 
Various methods have been proposed and used to solve this problem. 
For example, kinematic analysis of the longitude velocity diagram (LVD) assuming the Galactic rotation has been used to solve the line-of-sight degeneracy \cite{sofue1995a,sofue2017,oka+1998a,tsuboi+1999,kruijssen+2015,henshaw+2016,tokuyama+2019}.
Also, the method using absorption line profiles for background continuum emissions has offered powerful approaches \cite{sawada+2004,sofue2017,okatake+2022}.

With the LVD method several arm structures have been suggested.
The densest and most coherent LV ridge is called the Galactic Center (GC) Arm I, the second is Arm II, and further Arms (III and IV) have been proposed \cite{sofue1995a}.
More recent analyses have shown a larger number of arms with projected lengths $\sim 100-250$ pc \cite{henshaw+2016,tokuyama+2019}.

The association of radio continuum sources and HII regions with the molecular structures has been also an important subject related to the star formation (SF) activity in the GC \cite{sofue1990,henshaw+2022}.
The SF regions, Sgr B2 and C, are associated with the molecular complexes at the leading ends of Arm I and II, respectively \cite{hasegawa+1994,sofue1995a,sawada+2004}.

Despite of the extensive analyses about CMZ in the decades, however, a variety of conflicting models of the structure and kinematics have been proposed as is thoroughly summarized in a recent review article \cite{henshaw+2022}.
This is mainly because of the lack in precise observational information about the 3D distribution, motion, and mutual location of gaseous structures and radio sources. 

In this paper, we revisit this classical problem by analyzing data cubes of the CO and HI line emissions in the $(l,b,v)$ space. 
By confirming a tight correlation of the HI structures with CO, we use the HI absorption to investigate the line-of-sight (los) spatial relations of the molecular structures.
We further derive the 3D velocity vector of Sgr B2 with respect to Sgr A$^*$ using the data of trigonometric measurements of proper motions.
We also revisit the spatial correlation of radio continuum structures with the molecular gas distribution \cite{sofue1990} using the recent high-resolution radio imaging of the GC.
\red{In this paper, we adopt the GC distance of $R_0= 8.18$ kpc \cite{gravity+2019}.}

The purpose of the present paper is to impose observational constraints on the basic parameters of the 3D structure and kinematics of the CMZ by analyzing the most recent observational data taken from the literature and archives. 
The parameters will help to distinguish a reliable model from the variety of theoretical models of the GC.

\section{Data}
  
The CO-line, HI 21-cm line, and 1.28 GHz radio continuum data have been taken from the archives as follows.

The \thco data were taken from the CO-line survey of the central $3\deg\times 0\deg.7$ region using the Nobeyama 45-m telescope \cite{tokuyama+2019}, which had a FWHM (full width of half maximum) angular resolution of $15''$, velocity resolution of $1 \ekms$, rms (root mean squared) noise temperature $1$ K presented in a data cube with grid spacing of $(7''.5, 7''.5, 1 \ekms)$. 

The \twco data were taken from the survey of the central $\sim 4\deg \times 1\deg$ region using the 45 m telescope \cite{oka+1998a}, which had an effective angular resolution of $37''$ ($={\rm (beam~width^2+grid~spacing^2})^{1/2}$, velocity resolution $2$ \kms, rms noise of $\sim 1$ K, presented by a data cube with grid spacing of $(34'',34'',2 \ekms)$. 

The HI data were taken from the survey with the Australia Telescope Compact Array (ATCA) of the central $10\deg\times 10\deg$ region \cite{mcclure+2012}, which had an angular resolution of $145''$, velocity resolution of $1 \ekms$, rms noise of 0.1 K, and are presented in a cube of spacing $(35'',35'', 1 \ekms)$. 

While a variety of mm-wave lines have been used to analyze the molecular structures in the CMZ, detailed comparative study of the HI- and CO-line structures has not been obtained yet.
This is mainly because of the HI angular resolutions not sufficient to resolve the GC structures by the current single dish observations
\cite{kalberla+2005,HI4PI2016,mcclure+2009}.
From the CO and HI line data we construct integrated intensity (moment 0) maps to examine the spatial distributions of the molecular and atomic gases projected on the sky.
We also perform a detailed comparison of the HI and CO-line LVDs to investigate the kinematics of the gaseous structures. 

The CO line data are also compared with the radio continuum map in order to examine the spatial relationship on the sky of the molecular arms with star-forming regions.
The radio continuum map at 1.28 GHz was taken from the MeerKAT Galactic Center archive \cite{heywood+2022}, which had angular resolution of $4''$ after Gaussian convolution and grid spacing of $1''.1$.

\section{CMZ enveloped by Central HI Zone (CHZ)} 

\subsection{{Disc-eliminated intensity maps}}

{Because of the degeneracy of radial velocity near the $Y$ axis (Sun-GC line) as well as for the long line-of-sight depth from the Sun to the far-side disc of the Galaxy, integrated intensity maps of the CO and HI line emissions in the GC region are significantly contaminated by the fore- and background disc.
In order to avoid the contamination, we apply a method of Disc Elimination by radial Velocity cutting as described in the Appendix.
Considering the velocity dispersion and radial motions by spiral arms as well as the expanding rings like the 3-kpc arm, we take the cutting velocity to be $\vcut=50 \ekms$, so that the emission at $\absv \le \vcut$ is subtracted from the total integrated intensity maps.
Thereby, the emission from a fan-shaped region near the $Y$ axis in the GC disc itself is also eliminated.
This over-eliminated component is recovered by considering the proportion of the fan-shaped area to the entire disc (see Appendix).}

\begin{figure*} 
\begin{center}   
\includegraphics[width=17cm]{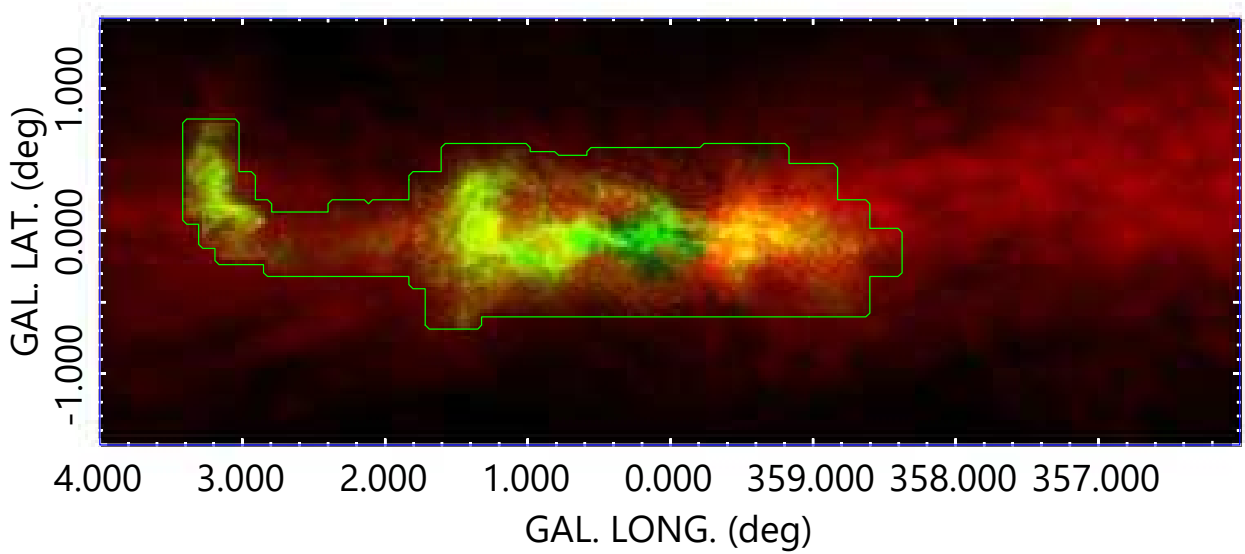} 
\end{center}
\caption{  
Superposition of the \twco map in green color on the HI map in  red color. Green box encloses the area of the Nobeyama CO survey. Disc elimination at $|\vlsr|\le 50$ \kms has been applied in both maps.
} 
\label{RG}
\end{figure*}

Fig. \ref{maps} shows the thus obtained integrated intensity (moment 0) maps using the \twco-line data \cite{oka+1998a} from the Nobeyama survey and HI from ATC survey \cite{mcclure+2012}.
Panel (a) shows the integrated intensity map over the entire spectral (velocity) range, which we call hereafter Map A (All velocity), (b) integrated in $\absv \le \vcut=50 \ekms$ as Map B (Background disc), and (c) shows the intensity integrated at $\absv > 50$ \kms as Map C (Center).
Namely, 
\be 
\rm Map\ C = Map\ A - Map\ B.
\ee
The lower panels (d), (e) and (f) show Maps A, B and C in HI, respectively.  
In Fig. \ref{RG} we present superposition of the DEV maps of the HI line emission in red color scaling (0 to 3000 K \kms) and \twco-line in green (0 to 2000 K \kms).  

In other words, 
Map A is the usual integrated intensity map; 
Map B shows intensity of the fore- and background disc (hereafter, disc) near the degenerated velocities around $l\sim 0\deg$; 
and Map C is the map showing intensity after subtracting the fore- and background disc.

Since we cut the emission at $|\vlsr|\le\vcut= 50\ekms$, the procedure also eliminates the low-velocity emission from the GC itself.
Therefore, when Map C is used to measure the masses of CMZ and CHZ (central HI zone), the over-eliminated emissions must be recovered using the method described in the Appendix.
The DEV maps are also used to define the CMZ and CHZ by measuring the cross sections along and perpendicular to the Galactic plane.

\begin{figure*} 
\begin{center}   \includegraphics[width=18cm]{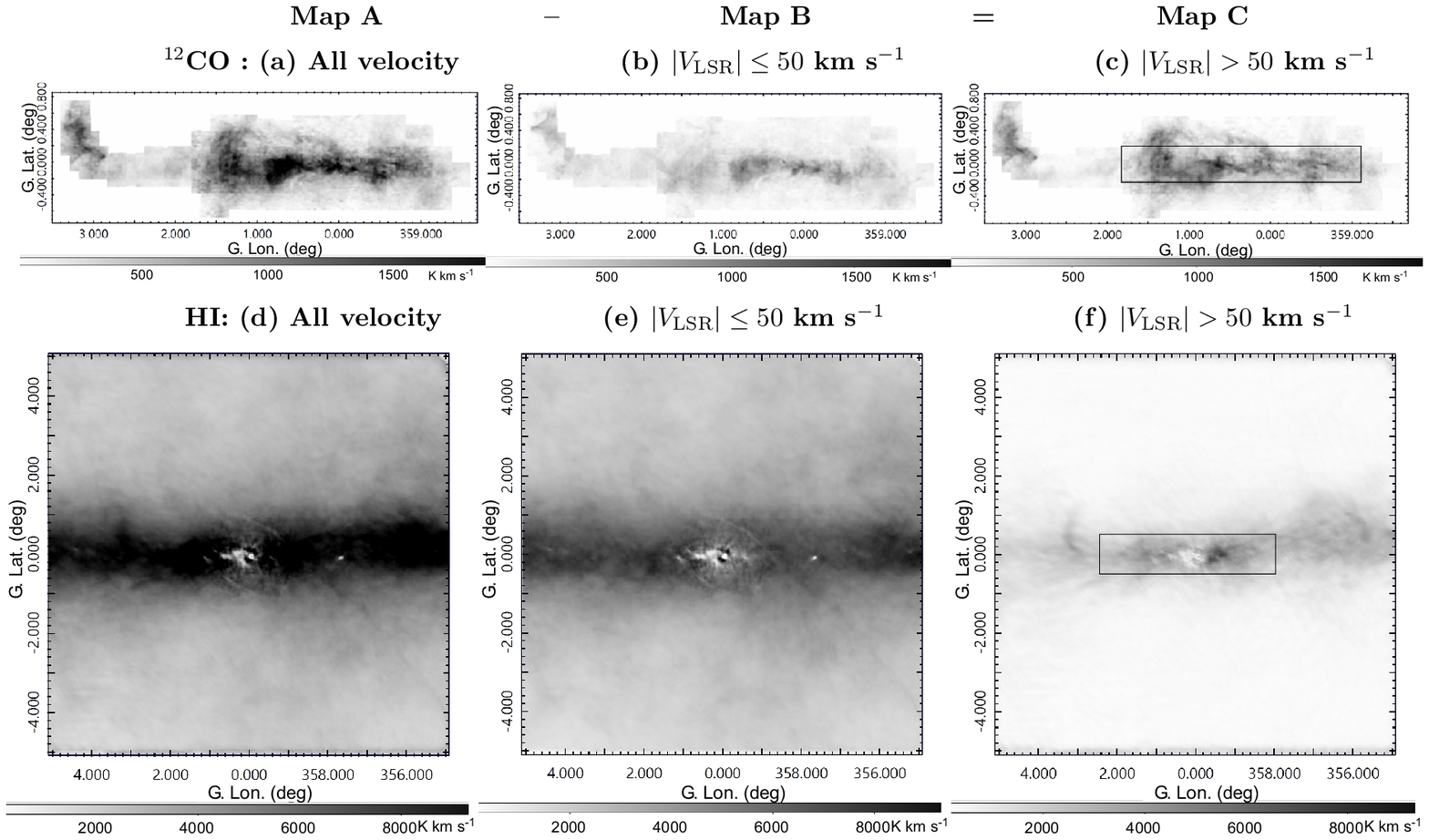}  
\end{center}
\caption{{[Upper panels] Integrated intensity ($\Ico$) maps of the \twco line emission from Oka et al. (1998). 
(a) Map A: Total $\Ico$ in All velocity range. Grey-scale bar indicates $\Ico$ in \Kkms.
(b) Map B: within $|\vlsr|\le 50$ \kms, representing fore- and Background disc emission.
(c) Map C: Outside $|\vlsr|>50$ \kms, representing GC emission. The squared box defines the CMZ.
[Bottom panels] Integrated intensity map of the HI line emission from McClure-Griffiths et al. (2012).  
(d) Map A for HI $\Ihi$. Intensity scale is \Kkms.  
(e) Map B, and (f) Map C. The squared box defines the CHZ.
} }
\label{maps}
\end{figure*}

\subsection{Definition of CMZ and CHZ from the intensity profiles} 
 
\begin{figure} 
\begin{center} 
\includegraphics[width=8cm] {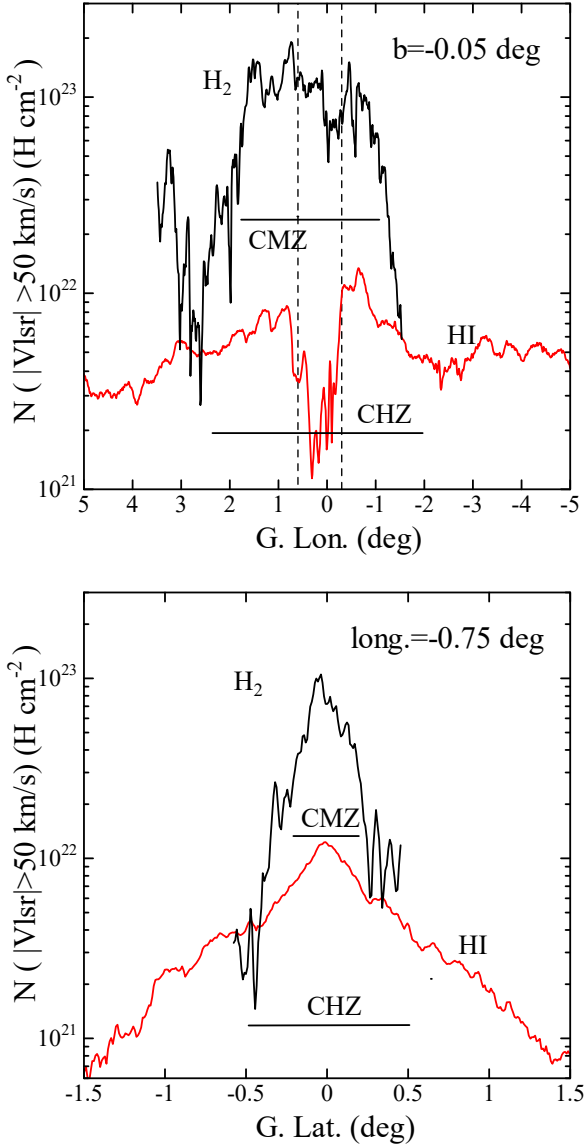}  
\end{center}
\caption{[Top] Horizontal cross sections at $b=-0\deg.05$ of hydrogen column density $N$ calculated from \twco and HI integrated intensities at $|\vlsr|\ge 50 \ekms$.
[Bottom] Same, but vertical cross sections at $l=-0\deg.75$. 
} 
\label{profile}
\end{figure}

In Fig. \ref {profile} we show longitudinal and latitudinal distributions of column densities $N$ of \Htwo and HI gases as obtained from the disc-eliminated intensity Maps C with $\vcut= 50 \ekms$.
The upper panel shows horizontal profiles at $b=-0\deg.05$, and the lower panel is a vertical profile at $l=-0\deg.75$.
The vertical dashed lines enclose the region where HI absorption against radio continuum sources is significant.
Here, we adopted the conversion factors as described in the next subsection.

The horizontal \Htwo (CO intensity) profile is plateau like with sharp shoulders at $l=-1\deg.1$ and $+1\deg.8$ ($-157$ and $+257$ pc) and full width is $w_r({\rm CO})=2\deg.9=414$ pc. The vertical profile is Gaussian like with the $e$-folding full width of $w_z=0\deg.4=57$ pc or a half width of $h_z ({\rm CO}) \sim 0\deg.2\ (\pm 29)$ pc from the peak position. 
We define the CMZ as the region enclosed by these shoulders as indicated by the thick lines in Fig. \ref{profile}.
We point out that the horizontal plateaued profile indicates a ring-like distribution of the molecular gas inside the CMZ. 
 
The HI profiles are milder, having no clear boundaries, and the profiles outside the absorption region near Sgr A and B can be approximated by an exponential function. 
The horizontal $e$-folding scale radius is measured to be $h_r\sim \pm 2\deg.2=314$ pc, and the vertical scale height to be $h_z\sim 0\deg.5=71$ pc.  
We define the central HI zone (CHZ) as this squared area of $w_r\times w_z({\rm HI})=4\deg.4\times 1\deg=629 \times 143$ pc as indicated by the horizontal lines in Fig. \ref{profile}.

\subsection{CMZ Mass}

The estimation of the mass of a "zone" in the GC direction is an ambiguous task by three reasons.
One is the area of integration that depends on the quality of data (sensitivity and resolution) and by the authors definition. 
Second is the contamination of the fore- and background Galactic disc. 
The third problem is the uncertain conversion factor from CO line intensity to \Htwo column density \cite{dahmen+1998}.

The first problem is solved by the new definition of the CMZ area in the previous subsection, and the second by the low-velocity cut method (DEV) for the line integration.
For applying the DEV, there appear three kinds of masses. 
\begin{itemize}
    \item $M_{\rm A}$ stands for the total mass using intensity integrated over all the velocity range.
    \item $M_{\rm C}^0$ is the mass after eliminating the fore- and background at $|\vlsr|\le \vcut$ (here 50 \kms).
    \item $M_{\rm C}$ is the final mass after correcting for the over-elimination of the near-$Y$ axis fan-region in $M_{\rm C}^0$.
\end{itemize} 

As to the conversion from \twco intensity to molecular column density, we use the relation
\be 
N_{\rm H_2}=\Xco \Ico,
\ee
where $\Xco$ is the CO-to-\Htwo conversion factor.
We adopt the empirical relation between $\Xco$ and GC distance $R$, considering the metallicity gradient in the Galaxy \cite{arimoto+1996}, and rewrite the formula in terms of the local value $X_0$ at $R=R_0$,
\be
\log \Xco(R)/X_0=0.41(R-R_0)/r_e. \label{logXco}
\ee  
Here, $r_e=6.2\times(R_0/10\ {\rm kpc})=5.1$ kpc is the scale radius of the metallicity gradient, which has been re-scaled to $R_0=8.18$ kpc from the original metallicity gradient of $0.07\pm 0.015$ dex per kpc derived for $R_0=10$ kpc \cite{shaver+1983}. 
Adopting the local value of  $X_0=\XCO(R_0)=2.0 \times 10^{20}$ \xcounit \cite{bolatto+2013,sofue+2020xco}, we obtain the conversion factor in the GC ($R=0$) to be
\be 
\Xco(0)=0.51\times 10^{20}\ {\rm H_2\ cm^{-2}\ [K\ km\ s^{-1}]^{-1}}.
\ee 

We first estimate the total luminosity without applying the disc elimination using Fig. \ref{maps} (a) by measuring the mean intensity in the defined CMZ area, $A$ in $l=-1\deg.1 =-157$ pc to $+1\deg.8 =+257$ pc and  $b=-0\deg.2=-29$ pc to  $+0\deg.2=+29$ pc, and obtain $\Ico\sim 1035$ \Kkms. 
The total molecular mass of CMZ is then calculated as
\be
M_{\rm A} ({\rm CMZ})=2\mu \mH A \Xco(0) \Ico \sim 2.8\times 10^7 \Msun.\label{mass1}
\ee 
Here, we assumed the mean molecular weight of $\mu=1.4$ per hydrogen atom including metals.

We then measure the mean \twco intensity in the CMZ on the disc-eliminated Map C in panel (c) of Fig. \ref{maps}, and obtain $\Ico\sim 681$ K \kms.
The disc-eliminated mass $M_{\rm C}^0$ of the CMZ is then estimated by multiplying the area of CMZ to be
\be
M_{\rm C}^0({\rm CMZ}) =1.8 \times 10^7 \Msun.\label{mass2}
\ee 

As described in the Appendix, we further correct for the over-elimination of the CMZ itself within the fan-shaped area of 
\be 
\phi =\sin^{-1} (\vcut/\Vrot)=19\deg.5
\ee 
for $\vcut=50$ and $\vrot=150 \ekms$.
This component is seen as an excess over the extended fore/background emission in Map B of panel (b) of Fig. \ref{maps}.
The fan correction factor is given by 
\be
\gamma=1/(1-2\phi/\pi)=1.28,
\ee 
and we obtain the corrected final mass of the CMZ by
\be
M_{\rm C}({\rm CMZ})=\gamma M_{\rm C}^0\sim 2.3 \times 10^7 \Msun. \label{mass3}
\ee

Comparison of $M_{\rm C}$ with the non-corrected mass, $M_{\rm A}$, tells us that the current estimation of the molecular mass in the GC has been systematically over-estimated by about 22\%. 

\begin{table*}   
    \caption{Molecular and HI masses in CMZ and CHZ$^\dagger$}
    \begin{tabular}{lllllll} 
    \hline 
Region & Mol. or HI & Area& Area& $M_{\rm A}$  & $M_{\rm C}$ &Method   \\ 
      && $l$ range & $b$ range& ($\Msun$) & ($\Msun$)  \\ 
    \hline 
CMZ & Mol. gas &$-1\deg.1\sim 1\deg.8$ &$-0\deg.2\sim+0\deg.2$ &$2.8 \times 10^7$ & $2.3
\times 10^7$& Sq. photom., area col. 2,3, Fig. \ref{maps}(a)(c)\\
ibid &HI&   ibid  & ibid  & $ 4.6 \times 10^6$& $2.3 \times 10^6$ & Sq. photom., area col. 2,3, Fig. \ref{maps}(d)(f)\\ 
\hline
CHZ &HI& $-2\deg \sim +2\deg.5$ & $-0\deg.5\sim +0\deg.5$& $1.8
\times 10^7$& $6.9 \times 10^6$ &Sq. photom., area col. 2,3, Fig. \ref{maps}(d)(f)\\ 
\hline 
    \end{tabular}\\
$^\dagger$ Conversion factor of $\Xco(0)=0.51 \times 10^{20}$ \xcounit is used. $M_{\rm A}$ stands for total integrated mass over all the velocity range.
$M_{\rm C}$ gives the mass after disc elimination by velocity at $|\vlsr|\le 50 \ekms$ and corrected for fan-shape over elimination. 
    \label{tabmass} 
\end{table*}

\subsection{CHZ Mass}

The HI mass of the CHZ is calculated by 
\be 
M=\mu \mH A \Xhi \Ihi,
\ee 
where
\be 
X_{\rm HI} =1.82 \times 10^{18} {\rm H\ cm^{-2} [K\ km\ s^{-1}]^{-1}}.
\ee 
This relation requires that the HI line is optically thin, which is approximately satisfied in the integrated velocity ranges, where the brightness temperature is mostly less than $\sim 40$ K.

The HI zone (CHZ) is extended without clear-cut boundaries. The longitude and latitude profiles in Fig. \ref{profile} indicate that the distribution can be approximately fitted by an exponential function of the longitude and latitude.
So, we measured the $e$-folding longitudes to be $l\sim -2\deg \ (286)$ pc and $\sim +2\deg.5$ (357 pc), and latitude heights are $b=-0\deg.5$ and $+0\deg.5$ ($\pm 71$ pc).

We also apply the disc elimination using the same method as for CMZ, and the obtained HI maps are shown in Fig. \ref{maps}. 
We use these maps to measure the integrated intensity by square-box photometry for the CHZ region as above as well as for the CMZ region given in the previous subsection.
We thus obtain the simply integrated mass
\be
M_{\rm A}({\rm CHZ})=1.8 \times10^7\Msun,
\ee
and the mass corrected for the disc elimination and over correction as
\be
M_{\rm C}({\rm CHZ})=6.9 \times10^6\Msun.
\ee
It is impressive that the total mass is 2.6 times greater than the disc eliminated and fan corrected mass.
This is because the disc emission dominates in the Galactic plane, and is much brighter than the GC emission.
For comparison, we also measure the corrected HI mass inside CMZ as
\be
M_{\rm C}({\rm HI\ in\ CMZ})=2.3 \times10^6\Msun,
\ee 
which is 10\% of the CMZ's molecular mass.
The results are listed in table \ref{tabmass}.

\subsection{Molecular fraction}

The molecular and HI masses in the same area of the CMZ yield an HI-to-\Htwo mass ratio of $\eta\sim 0.09$ and the molecular fraction of $\fmol =1/(1+\eta)\sim 0.9$. 
The molecular fraction can be calculated also locally using the column density plots in Fig. \ref{profile}, which varies from $\sim 0.9$ to 0.95 
inside the CMZ. 

The high molecular fraction in CMZ is consistent with that found in the GC region by analyses of the global distributions of HI and \Htwo gases observed in the Galaxy as well as the theoretical calculations assuming the exponential variation of the pressure $P$, metallicity $Z$, and ultra-violet radiation field $U$ \cite{sofue+2016,koda+2016}.
It is, however, significantly higher than the theoretical value of $\fmol\sim 0.5$ incorporating the phase transition between HI and \Htwo in the CMZ \cite{tress+2020}.
A detailed study of the molecular fraction in the GC will be presented in a separate paper \cite{sofue2022}.

We comment that the mass ratio between $M_{\rm C}({\rm CMZ})$ and $M_{\rm C}({\rm CHZ})$ is $\eta \sim 0.25$, which gives an apparently lower  molecular fraction $\fmol\sim 0.8$.
However, this milder value is due to the different areas (extent and height) of mass estimations for CHZ and CMZ, and hence, it does not represent the ISM condition. 

\subsection{Comment on the over-estimation of the mass in the GC}

The difference by a factor of 1.2 between the molecular mass of CMZ by all-velocity integration ($M_{\rm A}$)  and the DEV mass by low-velocity cutting after the correction for the over-eliminatin ($M_{\rm C}$) must be taken into account, when we discuss details of the mass distribution in the GC.
Namely, we must be aware that the current mass estimations of molecular gas in the GC have over-estimated it by about $\sim 20$\%. 
This comes mainly from the the 4-kpc molecular ring and its counter arm, and the number of intervening spiral arms in front and beyond the GC.   
Such an overestimation is much larger and far more crucial in the HI mass calculation, because the HI in the outer disc (near and far) is much more brighter than the CHZ. 

\section{GC Arms and Star-Forming Ring} 

\subsection{Arms I and II}

\begin{figure*}
\begin{center} 
\includegraphics[width=12cm]{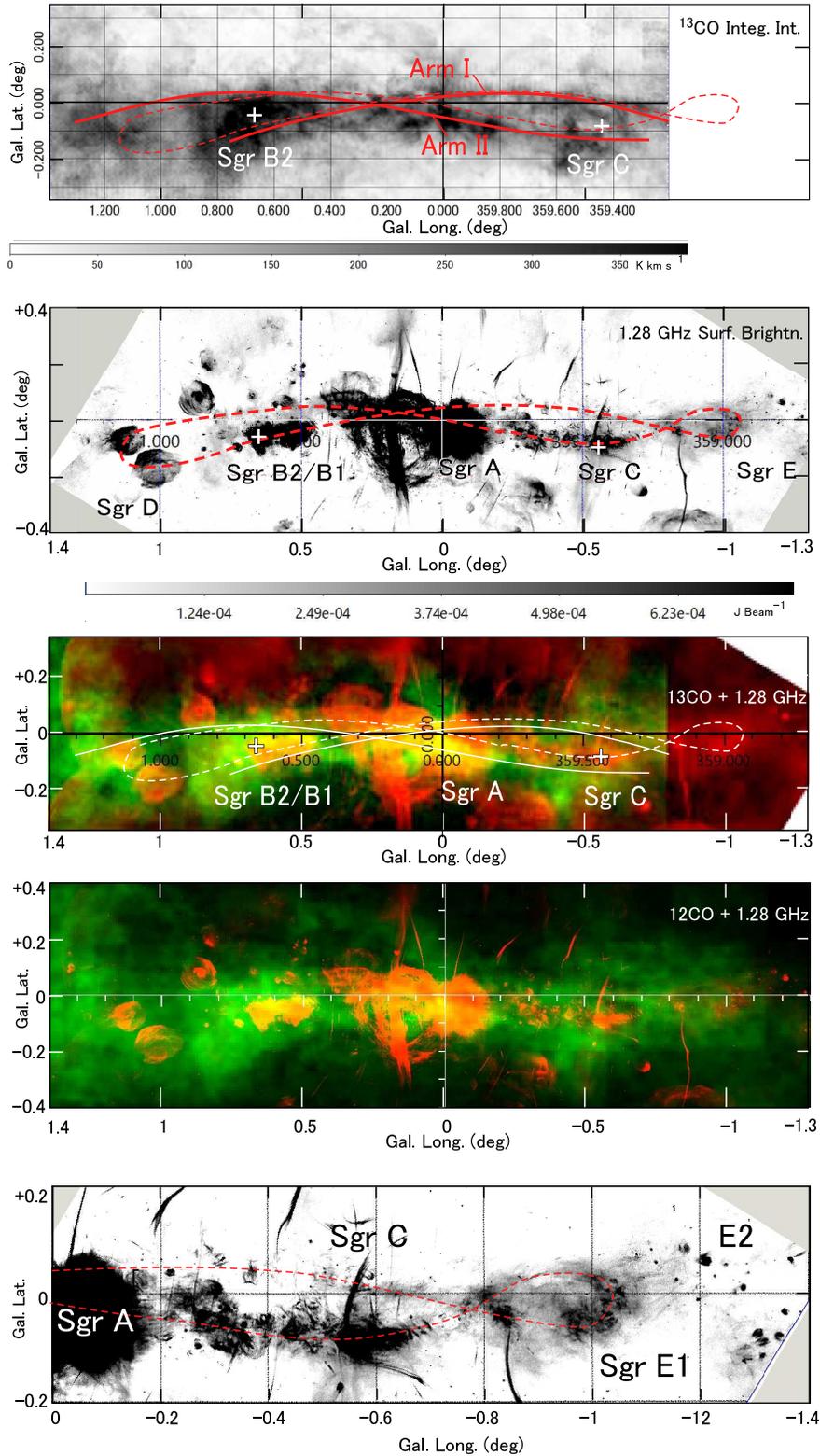} 
\end{center}
\caption{[Top] \thco-line integrated intensity (moment 0) map made from the CO survey with the Nobeyama 45-m telescope (Tokuyama et al. 2019).
GC Arms I and II are traced by the full lines. 
[2nd] MeerKAT image of the radio continuum emission at 1.28 GHz (Heywood et al. 2019; Yusef-Zadeh et al. 2022). 
The radio belt is traced by the dashed line, indicating a warped ring linking Sgr B1, B2, C, D and E.
[3rd] Superposition of the \thco intensity map (green) on 1.28 GHz map (red).
[4th] Same, but \twco map (green; from 0 to 2500 K \kms; Oka et al. 1998) on 1.28 GHz (red).
[Bottom] Enlargement of the radio belt near Sgr C and E from the 2nd panel.} 
\label{belt}
\end{figure*}

Fig. \ref{belt} (top panel) shows a moment 0 (integrated intensity, $\Ico$) map of the $^{13}$CO line emission integrated over the velocity range of $-220 \le V_{\rm LSR} \le 220$ \kms using the \thco-line survey \cite{tokuyama+2019}.
Here we used the \thco line, which is less contaminated by the disc than the \twco line, and is more appropriate to abstract arm and cloud structures in the CMZ. 

The two major molecular arms named as GC Arm (GCA) I and II \cite{sofue1995a} are traced by the red lines.
The arms compose coherently stretched ridges extending over $\sim 1\deg.5$ in the longitudinal direction, and are slightly curved and bent concave to the north.
The latitudinal full width of the arms is as narrow as $\sim 0\deg.2-0\deg.3$ ($\sim 29$ to 43 pc). 

Arm I is associated with the Sgr B molecular complex at $l\sim 0\deg.6$ to $\sim 0\deg.9$ nesting the HII regions Sgr B2 and B1.
Arm II is associated with a molecular clump at $l \sim -0\deg.4$ to $\sim -0\deg.6$, which nests Sgr C. 
The molecular complex and clump are locally extending in the latitude direction, reaching $b\sim -0\deg.2$. 
The total molecular mass of the Arms and complexes shares $\sim 80$\% of the total molecular mass of the CMZ \cite{sofue1995a}.
 
\subsection{Radio belt from Sgr E through D}

Comparison of the distributions of the CO-line intensity and 10 GHz radio continuum emission in the GC on the sky has shown their global mutual association as well as their local avoidance with each other in smaller scales \cite{sofue1990,hasegawa+1994}. 

We here compare in more detail the CO line intensity map from the Nobeyama 45-m telescope \cite{oka+1998a,oka+1998b,tokuyama+2019} with the high-resolution radio continuum map at 1.28 GHz from MeerKAT \cite{heywood+2022,yz+2022}, which are shown in Fig. \ref{maps} and \ref{belt}. 
Besides Sgr B, C, D and E, the molecular disc is shown to be associated with a horizontal belt of numerous HII regions with thermal radio emission having flat radio spectra \cite{yz+2022}. 

We first notice a prominent belt (or chain) of extended radio sources running through Sgr C in the 2nd panel as marked by the dashed line, which apparently starts from $(l,b)\sim(-0\deg.2,-0\deg.04)$ and runs across Sgr C.
The belt runs nearly parallel to the molecular Arm II shown by the red line in the top panel.
The radio sources along the belt have flat radio spectra indicating that they are mostly thermal objects \cite{yz+2022}.

The belt further extends to the west, crossing the Galactic plane around $l\sim -0\deg.8$, and reaches $l\sim -1\deg.1$ at the eastern clump of HII regions in Sgr E, which we name here Sgr E1. 
The radio clump E1 appears to be a separate structure from the cluster of HII regions distributed along the western-most dust ridge \cite{anderson+2020}, which we call here Sgr E2.
The bottom panel of the figure enlarges the radio belt, and the major radio continuum sources.

The radio belt, then, bends sharply in Sgr E1 at $l=-1\deg.1$, and returns to the east by drawing an $\infty$ shape, while the returning belt is weaker.
It then crosses again the Galactic plane at $l\sim -0.8$, and extends to the east parallel to the molecular Arm I. 
It is once disrupted by Sgr A around $l\sim -0\deg.2$, but again continues to positive longitude side parallel to Arm I.
The belt reaches Sgr B1 and B2, which is enveloped by an extended radio emission elongated toward the west, and can be linked to Sgr D.

Symmetrically to Arm I's radio belt,  the belt coming from Sgr C  along Arm II is traced toward the east, running near SNR G0.9+0.1, and reaches Sgr D.
There, it is connected to the belt coming from the south along Arm I to compose a continuous bent belt as marked by the dashed line. 

\subsection{Sky Projection of the Star-Forming Ring Drawing a Double Infinity }

We have thus linked the thermal radio sources (HII regions Sgr B1, B2, C, and E1) and Sgr D by a continuous belt as indicated by the dashed line in Fig. \ref{belt}. It draws a double-infinity ($\dinfty$) shape on the sky with the vertical amplitude of $\delta z \sim \pm 0\deg.06$
($\pm 8.6$ pc), and attains three times the maximum $z$ heights.
The particular shape suggests that the belt is an edge-on projection of a warping/waving ring of star forming regions in accordance with the bending Arms I and II of molecular gas that compose a warped ring tilted by $5\deg$ from the Galactic plane \cite{sofue1995a}.
This view is also in accordance with the warped ring exhibiting an infinity-shape \cite{molinari+2011,henshaw+2016}. 

Thus, the present double infinity has expanded the GC ring from the area at $l\sim \pm 0\deg.8$ between Sgr B and C to a larger and more perfect ring with radius $l\sim \pm 1\deg.2$, and links the major four radio sources including Sgr D and E.
We also point out that the double infinity has three maximum heights in the $z$ direction.
It has no fixed rotation axis that defines a tilted disc or a ring.
So, the wavy nature of the $\dinfty$ has no direct correlation with the larger scale HI disc with a tilt angle $\sim 10-20\deg$ \cite{liszt+1980,mcclure+2012,krish+2020}, the expanding molecular cylinder with axis at $PA\sim -15\deg$ tilted toward west \cite{sofue2017}, or the extended hot plasma with minor axis direction at $PA\sim -20\deg$ \cite{yamauchi+1990}. 

The $\dinfty$ shape can be interpreted as due to an oscillation of a rotating ring in the vertical direction \cite{tress+2020}, which has three amplitude maxima, and hence three times oscillation in one rotation.  
Since Arm I is located in the near side and rotation is clockwise as seen from the North Galactic Pole, as discussed in later section, the ring has three ascending nodes, which appear at $l=-0\deg.7, \ +0\deg.8$ and at $-0\deg.9$.
The oscillation period is one third of the rotation period, which is on the order of $P_z \sim (1/3) 2\pi R/\vrot \sim 1.2\times 10^6$ y for $R\sim 120$ pc and $\vrot\sim 200$ \kms.  

In the 3rd and 4th panels of Fig. \ref{belt} we superpose integrated intensity (moment 0) maps of the \thco \cite{tokuyama+2019} and \twco line \cite{oka+1998a} in green color, respectively, on the 1.28 GHz radio continuum map in red.
The molecular Arms are associated with the radio belt as indicated by the full and dashed lines, respectively.

It is interesting to point out that the molecular Arms are vertically displaced from the radio belt.
The local displacement of HII regions and molecular clouds, and the systematic displacement between molecular and SF spiral arms seen from the Galactic poles, are well known and explained in terms of star formation mechanisms in density-wave and galactic shock theories of spiral arms.
However, the systematic displacement of the molecular and radio bands perpendicular to the galactic plane is noted here for the first time, raising new questions about the vertical bias of SF activity in the GC disk.

\subsection{Sgr D and SNRs}

Although Sgr D has been suggested to be fore- or background objects independent of the GC \cite{kauffmann+2017}, we reconsider here its distance and association with the GC based on the recent high resolution radio continuum data \cite{yz+2022,henshaw+2022}.
Sgr D is composed of three SNRs and two HII regions \cite{mehringer+1998}.
The easternmost radio shell G1.1-0.1 has a steep radio spectrum of $\alpha \sim -1$  indicative of a SNR \cite{yz+2022}.
The shell is overlapped by an HII region G1.15-0.07, which is a foreground object at a distance of 2.36 kpc from the Sun according to the trigonometric measurements by VERA \cite{sakai+2017}.
G1.0-0.2 is a typical shell-type SNR with nonthermal radio emission.
G0.9+0.1 is a nonthermal radio shell also overlapped by a bright HII region with brightness temperature of $\Tb\sim 270$ K  \cite{yz+2022}. 

In order to examine if these shell-type SNRs are GC objects, we estimate their distances using the $\Sigma-D$ 
(surface brightness-diameter) relation at 1 GHz
empirically fitted to SNRs with known distances including those of the LMC and SMC,
\be
\Sigma_{\rm  1GHz}=2.07^{+3.10}_{-1.24}\times 10^{-17} D({\rm pc})^{-2.38\pm0.26}{\rm w~ m^{-2} Hz^{-1} str^{-1}}
\ee
\cite{case+1998}. 
Mean surface brightness at 1.28 GHz was measured on the MeerKAT image after subtracting the HII regions, and converted to brightness at 1 GHz assuming a radio spectrum of $\alpha = -1$. 

The estimated linear diameters and distances from the Sun are listed in table \ref{tabSNR} along with the mean brightness and angular sizes.  
Considering the scatter and uncertainty of the $\Sigma-D$ relation, which is on the order of $\pm 30$\%, we may conclude that the three SNRs are GC objects.
Their diameter $\sim 20$ pc indicates a young age of $t\sim 10^3-10^4$ y, suggesting that the Sgr D area was a recent SF site.
Also recalling that the region from $R\sim 0.3$ to $\sim 2$ kpc around the GC is almost empty in molecular clouds and SF sites, we may consider that the distance coincidence around 8 kpc and the young ages strongly support that the SNRs are GC objects.
So, we here assume that they are associated with the CMZ.

\begin{table} 
    \centering
    \caption{SNRs in Sgr D region}
    \begin{tabular}{lllll} 
    \hline 
    SNR & $\Sigma_{\rm 1GHz}$  &Ang. dia  & Diam.  & Dist.  \\ 
      &  {\tiny (w m$^{-2}$Hz$^{-1}$str$^{-1}$)} & (arc min) &   (pc) &  (kpc)\\ 
    \hline 
G1.1-0.1&    1.84&    7.4&   19.2&    8.9\\
G1.0-0.2&   1.47&    9.1&   21.1&    8.0\\
G0.9+0.1&    1.42&    7.6&   21.4&    9.6\\
\hline 
    \end{tabular}
    \label{tabSNR}
\end{table}

\section{Kinematics and 3D Structure}

\subsection{Longitude-Velocity Diagrams (LVD): Tight correlation of CO and HI features}

\begin{figure} 
\begin{center} 
\includegraphics[width=8cm]{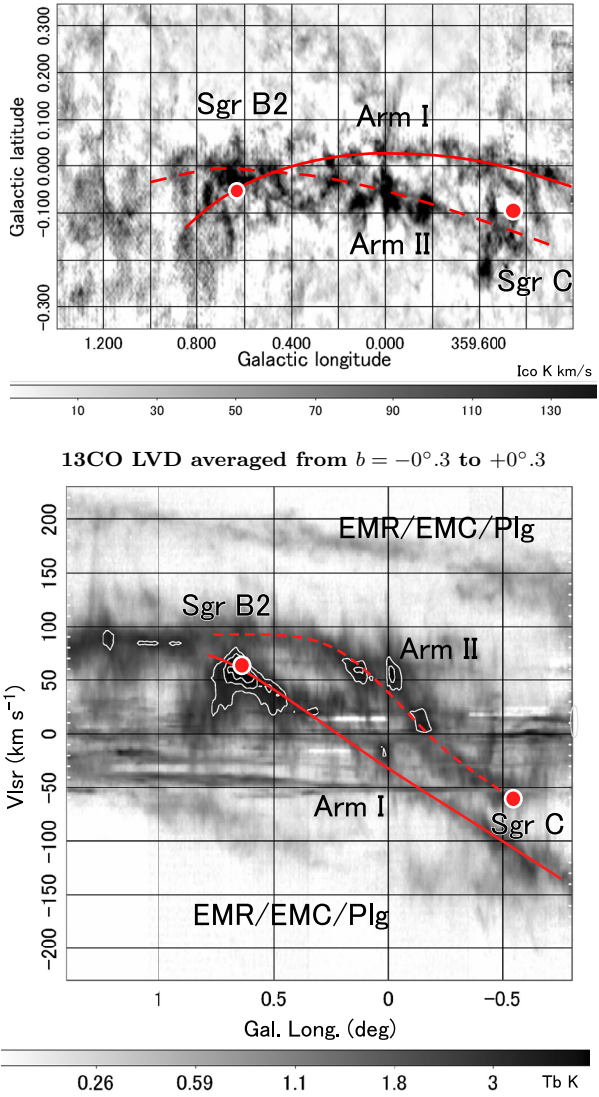}
\end{center}
\caption{{[Top] Background-filtered map of integrated intensity of \thco emission with the $b$ (vertical) axis enlarged twice, and [bottom] LVD of \thco line emission from the Nobeyama Bears CO survey (Tokyuama et al. 2019). Major structures discussed in the text and  Fig. \ref{LV-RG} are labelled. Intensity scale in LVD is logarithmic. Note that EMR/EMC/Plg is an order of magnitude less bright than Arms.
}}
\label{lv13co} 
\end{figure}

Since the CMZ and CHZ as well as the radio continuum belt are sky projections of edge-on disc and arms, their 3D structures are degenerated in the intensity maps.
Longitude-velocity diagram (LVD) is a powerful tool to solve this degeneration.
Fig. \ref{lv13co} shows an LVD of the \thco line emission in comparison with the integrated-intensity map as reproduced from the Nobeyama CO-line survey \cite{tokuyama+2019}.
Extended structures in the intensity map has been subtracted using the background-filtering technique, and the vertical scale is enlarged twice, so that Arms I and II are more clearly visible than in the \twco map in Fig. \ref{maps}.
The LVD has been averaged between $b=-0\deg.3$ and $+0\deg.3$ , where the major structures discussed in this paper are labelled.
Arms I and II compose the major structure in the CMZ drawing tilted straight LV ridges across the coordinate origin. 

The expanding molecular ring (EMR) and cylinder (EMC) \cite{sofue1995b,sofue2017}, or the \red{parallelogram in the bar model \cite{fux1999,sormani+2018,henshaw+2022}}, show up as a diffuse features at high forbidden velocities, labelled as EMR/EMC/Plg.
Approximate positions of Sgr B and C are also marked on the eastern end of Arm I and western end of Arm II, respectively. 

In order to see how the HI gas distribution is correlated with the molecular structures, we then compare the CO LVDs with HI.
Fig. \ref{lv13co}(b) shows an $\Tb$ LVD of the \twco in green color and HI in red at a latitude of $b=+0\deg.05$.
Panel (c) shows the same but at $-0\deg.05$ and (d) at $-0\deg.3$.  
From these superposed maps, we can recognize the tight correlation of the CO and HI structures in the GC. 

\begin{figure*}  
\begin{center}
\vskip -3mm
\includegraphics[width=8.5cm]{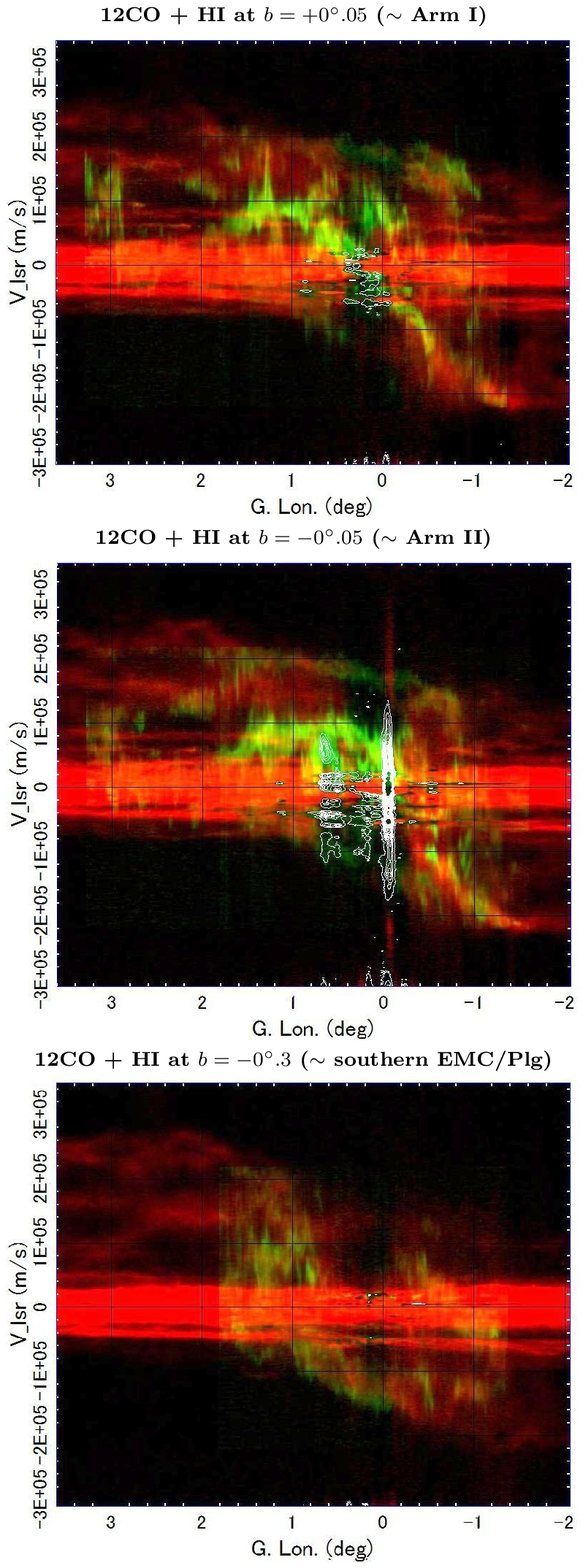} 
\end{center}
\vskip -3mm\caption{HI LVD ($\Tb$ in red from 0 to 50 K) superposed with $^{12}$CO LVD (green from 0 to 20 K). White contours are negative HI $\Tb$ from $-10$ K at interval $-50$ K. (HI: McClure-Griffith et al. 2012; CO: Oka et al. 1998).
[Top] At $b=+0\deg.05$, approximately along Arm I.
[Middle] At $b=-0\deg.05$,  partially along Arm II.
[Bottom] At $b=-0\deg.3$, showing round LV of EMR/EMC/Plg. Molecular Arms I and II are not visible at this height.
} 
\label{LV-RG} 
\end{figure*}

\subsection{GC Arms I and II in the LVD}

It is stressed that the molecular Arms I and II in the CO line are also well traced in the HI LVD as shown in Fig. \ref{LV-RG}.
Arm I is traced as the most prominent and longest ridge both in CO and HI, starting at $(l,\vlsr)\sim (0\deg.66,+70 \ekms)$ near Sgr B2 and extending straightly to the bottom-right corner of the diagram.
The HI LV ridge extends further toward negative longitudes, reaching $(l,\vlsr)\sim (-2\deg,-200 \ekms)$.
The main CO ridge intersects the rotation axis at $(l,\vlsr)=(0\deg,-30 \ekms)$ corresponding to a displacement of the apparent rotation center to $l\sim +0\deg.2$. 

Arm II is also traced as an LV ridge in CO and HI, starting from the left-side edge of the top panel at $(l,\vlsr)\sim (+1\deg.5, 100 \ekms)$, keeping the velocity till $(0\deg.15)$.
There, it bends and further extends, making a straight tilted LV ridge, and reaches $(-0\deg.7, -140 \ekms)$ near Sgr C. 
The main CO ridge intersects the rotation axis at $(l,\vlsr)=(0\deg,+40 \ekms)$ with apparent rotation center at $\delta l\sim -0\deg.15$.

The straight LV ridges along Arms I and II, nesting Sgr B and C, can be interpreted as due either to a rotating solid disc or a ring.
Because the former case (solid disc) is not allowed in the gravitational potential in the GC, it can be attributed to a rotating ring.
\red{Largely elongated elliptical orbit in a bar that postulates velocity splitting greater than $\sim \pm 30-40 \ekms$ at $l\sim 0\deg$ cannot explain the straight LV ridge.}

\subsection{Expanding Ring/Cylinder/Parallelogram}

The expanding molecular ring/cylinder/parallelogram (EMR/EMC/Plg) is visible at high non-circular velocities, extending in the latitude direction and composing a cylindrical structure.
The EMC mass has been estimated in our earlier work \cite{sofue1990b,sofue1995b,sofue2017}, and we re-scale it using the presently adopted conversion factor, obtaining
$M_{\rm EMC} \sim 4.5 \times 10^6 \Msun$.
So, the EMC's mass is only about $\sim10-20$\% of the CMZ.
 
The EMC is visible both in the CO and HI LVDs as oval ridges at positive and negative high velocities at $\vlsr \pm 100 - 200 \ekms$.
The LV ellipse is clearer at higher latitudes at $|b|\sim >0\deg.3$ ($>40$ pc). 
The $^{12}$CO LVD of EMR/EMC can be fitted by a cylinder with radius $\sim 200$ pc, expanding at $v_{\rm exp}\sim 160 \ekms$ with rotation speed of $V_{\rm rot}\sim 75$ \kms \cite{sofue2017}.

It is interesting to point out that the EMC's LV feature is fainter or even lacking near the Galactic plane.
This suggests that the cylinder is swept away or broken by the interaction with the CMZ Arms.

\subsection{3D kinematics from proper motion}
 
We determine the 3D velocity of Sgr B2 with respect to Sgr A$^*$ using the proper motions of Sgr A$^*$ and maser sources in Sgr B2, which have been measured extensively in the decades using the VLBA (Very Long Baseline Array) 
\cite{reid+2009,reid+2020},
and VERA (VLBI Exploration of Radio Astrometry)
\cite{oyama+2021,vera+2020}.
Combining the proper motions with the radial velocity, we can calculate the 3D velocity vector of Sgr B2 with respect to Sgr A$^*$.
In table \ref{tab1} we list the observed values, and calculated results. 

Taking an average of the measurements, we obtain the 3D velocity vector as follows. 
Sgr B2 is moving toward increasing longitude (to east) at $V_l\simeq 87.4$ \kms with respect to Sgr A$^*$, 
vertically toward negative latitude at $V_b\simeq -28.3$ \kms, 
and at radial velocity of $\vlsr=+62.3 \ekms$ in the line-of-sight direction.
These values yield rotational velocity of 107.3 \kms projected on the Galactic plane, and absolute velocity $V\simeq 111.0$ \kms. 

If we assume a circular orbit, the velocity vector can be used to determine the spatial position of Sgr B2 with respect to Sgr A$^*$ by combining with the projected distance on the sky of $0\deg.66$ (94 pc), which yields a galacto-centric distance of $r=160$ pc from Sgr A$^*$ at a position angle of $\theta=144\deg.9$ as seen from the North Galactic Pole, or making an angle $36\deg$ to the line of sight.
 
From the velocity vector, the inclination angle of the orbit is determined to be $i=-14\deg.2$.
This yields a projected tilt angle of the motion on the sky of $p=-18\deg.0$ from the Galactic plane, or Sgr B2 is moving toward the south west at a position angle of $PA=108\deg$ in the Galactic coordinates.

We comment that such a tilted orbit may be related to the curved ridge of Arm I as shown in Fig. \ref{belt} and to the vertical oscillation of the warped $\dinfty$ SF ring. 
Sgr B2 is located at the left-side end of the Arm I, significantly below the galactic plane, apparently leading the Arm toward PA$\sim 108\deg$.
Such an inclined orbit, or perpendicular motion in the rotating frame, may be related to the vertically extended structure of the Sgr B2 molecular cloud.

\begin{table*}
\caption{Proper motion and relative velocity of Sgr B2 about Sgr A$^*$.}
\begin{tabular}{llllllllll}
\hline\hline
Object 
& $\mu_\alpha\cos \delta$ 
&$\mu_\delta$
&$\mu_l \cos b$
&$\mu_b$
& $\mu_{l,{\rm BA}}$
&$\mu_{b,{\rm BA}}$
& $V_{l,{\rm BA}}$
&$V_{b,{\rm BA}}$
& $\vlsr$\\
~~
&(mas y$^{-1}$)
&&&&&
&(\kms)&&
 \\
\hline

VLBA$^1$ \\


Sgr A$^*$ 
&&
&{\tiny $-6.411\pm 0.008$} 
&{\tiny $-0.219\pm 0.007$}  
&&&\\

Sgr B2N 
&{\tiny $-0.32\pm 0.05$} 
&{\tiny $-4.69 \pm 0.11$}
&$ -4.28\pm 0.10$
&$ -1.95\pm 0.07$
&$ 2.13\pm 0.10$
&$ -1.73\pm 0.07$
&$ 82.7\pm 3.9$
&$ 67.1 \pm 2.7$
&$+64\pm 5$  \\

Sgr B2M 
& {\tiny $-1.23 \pm 0.04$ }
& {\tiny $ -3.84 \pm 0.11$ }
& $-3.96 \pm   0.10$
& $-0.74\pm   0.06$
& $2.45\pm  0.10$
& $-0.53\pm   0.06$
& $94.9 \pm 3.9$
& $ -20.4 \pm   2.5$
& $+61 \pm 5$ \\


\hline

VERA$^2$\\
Sgr A$^*$ 
&&
& {\tiny $-6.307 \pm 0.025$}
& {\tiny $-0.214 \pm 0.017$}
&&&\\


Sgr B2
& {\tiny $-1.83 \pm 0.21$ }
& {\tiny $-3.70 \pm 0.09$ }
& $-4.13\pm      0.14$
&$-0.15\pm  0.19$
& $2.18\pm 0.13$
&$  0.065\pm   0.19$
&$  84.7 \pm 5.0 $
&$  2.6 \pm  7.4$
& $+62.0 \pm 5.0 $ \\


\hline

Average$^\dagger$
&& &&
& $2.25\pm 0.17$
& $-0.73\pm 0.91$
& $87.4 \pm 6.5$
& $-28 \pm 35.5$

&62.3\\
\hline\\
\end{tabular}
\\
Data by small characters are from the literature. 1 \cite{reid+2020,reid+2009};
2 \cite{oyama+2021,vera+2020}. Suffix "BA" stands for values of Sgr B2 relative to A$^*$. Velocities were calculated for $R_0=\red{8.18}$ kpc.
\\
$^\dagger$ Average of the three measurements. Errors are standard deviations around simple means for equal weighting. \\
\label{tab1}
\end{table*}

\subsection{Arm's curvature by {\it dv/dl} method}

We try to estimate the curvature of Arms I and II by assuming that they are rotating around the GC.
If rigid-body tilt of the LVD is attributed to circular (nearly circular) motion, the velocity gradient $dv/dl=d\vlsr/dl$ of the ridge is related to the curvature  $r_{\rm curv}$ of the arm/ring through \cite{sofue2006}
\revone{
\begin{equation}
r_{\rm curv} \simeq R_0 V \left( \frac{d v}{dl}\right)^{-1} . 
\end{equation}
}
Using the LVD, we measured the gradient around $l\sim 0\deg$ to be
$dv/dl\sim 147 $ \kms deg$^{-1}$ for Arm I, and $\sim 193$ \kms deg$^{-1}$ for II. 
So, if we assume the same rotation velocity as Sgr B2 of $V\sim 106$ \kms, we obtain the approximate curvature of Arm I to be $r_{\rm curv, Arm I}\simeq 103$ pc, and Arm II $r_{\rm curv, Arm II}\simeq 79$ pc. 
Note that the method applies even to an oval or elliptical ring, but it does not tell us in which side the arm is located, near or far from the center.
 
The curvature $r_{\rm curv, Arm I}\simeq 103$ pc of Arm I is less than the GC distance of Sgr B2, $R_{\rm Sgr B2}\simeq 160$ pc, as derived from the geometry of 3D velocity vector on the assumption of circular rotation. This discrepancy may be resolved by the following possibilities: 

i) If we adopt 160 pc as the GC distance of Sgr B2 and assume that Arm I is not strongly deviated from an oval shape, the smaller curvature at the transverse point (TP) on the Sun-GC line may be understood as due to a bend of Arm I near TP. 
This means that the orbit is slightly elongated in the line of sight direction with its periphery near the TP. 

 ii) GC distance of T point is equal to $r_{\rm curv}$ and the orbit is highly elliptical with Sgr B2 having non-transverse motion with respect to Sgr B2-GC line.
Then, Sgr B2 is located near the periphery at closer distance to GC with slower velocity than the circular velocity in the gravitational potential. 
{However, such an elliptical orbit requires the gas to be oscillating almost radially, and hence, the straight LV ridge is not explained.}

iii) Both $R_{\rm Sgr B2}$ and $r_{\rm curv}$ are not directly related to the distance from the GC.
The data simply tell only the 3D velocity of Sgr B2 and Arm I's curvature at TP.
This means that we cannot argue about the orbits of the Arms and Sgr B2 or C, and leaves large freedom to fitting procedure by simulations, although the answer may be between i) and ii). 

As done in the previous subsection, we here rely on possibility i), and try to draw the 3D orbits of the Arms and Sgr B2 and C.
    
\subsection{HI Absorption Profiles of Sgr A to D: Line-of-sight locations}

\sss{HI Absorption spectra}

The fact that the CO arms are closely associated with HI makes it possible to examine the line-of-sight (los) location of the arms with respect to radio sources from the HI line absorption against continuum emission.
HI absorption spectra and los position determination have been extensively obtained  using the VLA toward radio sources and filaments in the GC \cite{lang+2010}.
We here use the HI data from the ATCA HI survey \cite{mcclure+2012} and examine the HI absorption toward the radio sources on Arms I and II.

\begin{figure} 
\begin{center}    
\includegraphics[width=8cm]{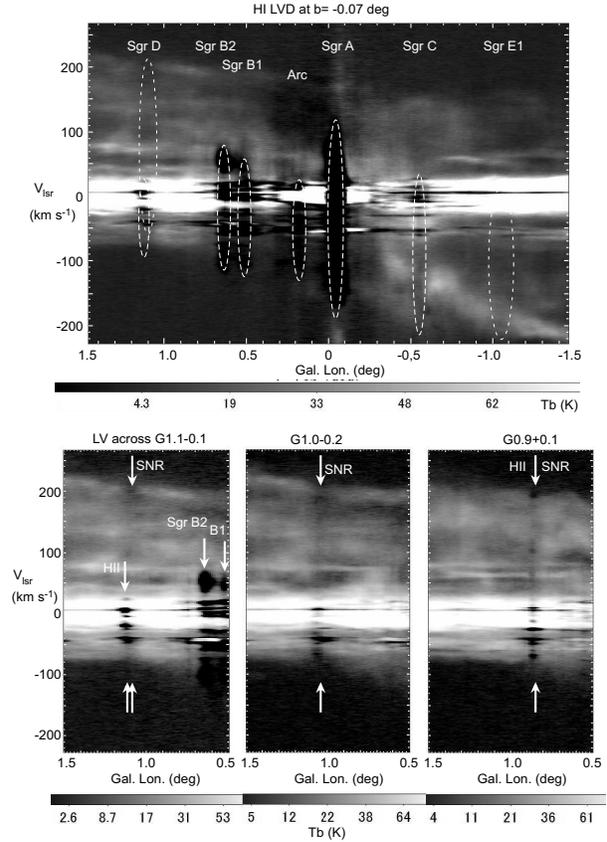}
\end{center}
\caption{HI LV diagrams at $b=-0\deg.07$ and across the SNRS and HII regions in Sgr D region, showing HI absorption against radio continuum sources. }
\label{lv-absorption}  
\end{figure}

\begin{figure} 
\begin{center} 
 \includegraphics[width=8cm]{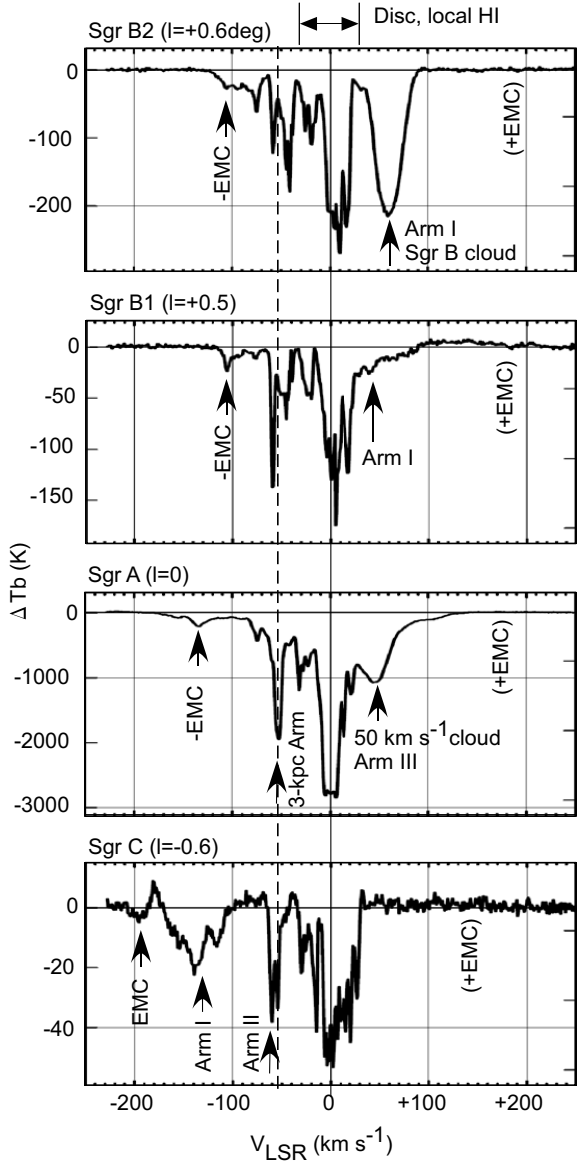}
 \end{center}
\caption{HI absorption spectra of $\Delta \Tb$ toward radio continuum sources Sgr B1, B2, A, and C (in the order of decreasing longitude) made by on-source minus  off-source spectra.}
\label{HIabsspec} 
\end{figure}

Fig. \ref{lv-absorption} shows an LVD at $b=-0.\deg.07$ across the major radio sources from Sgr A to E, where are marked some absorption features by dashed ellipses.
Each absorption spectrum, $\Delta \Tb$, toward a radio source was obtained by subtracting an average of off-source HI profiles in both sides of the source about one beam size.
The obtained spectrum represents the optical depth of the foreground HI gas in terms of the brightness temperature of the background continuum source $T_{\rm c}$.
$\Delta \Tb=-T_{\rm c} (1-e^{-\tau})$.
Fig. \ref{HIabsspec} shows the thus obtained HI absorption spectra toward Sgr B2, B1, A, and C in the order of decreasing longitude. 
They are almost identical with those obtained by the VLA observations \cite{lang+2010}, while slight differences are due to the difference in the angular resolutions.

\begin{figure} 
\begin{center}
\includegraphics[width=8cm]{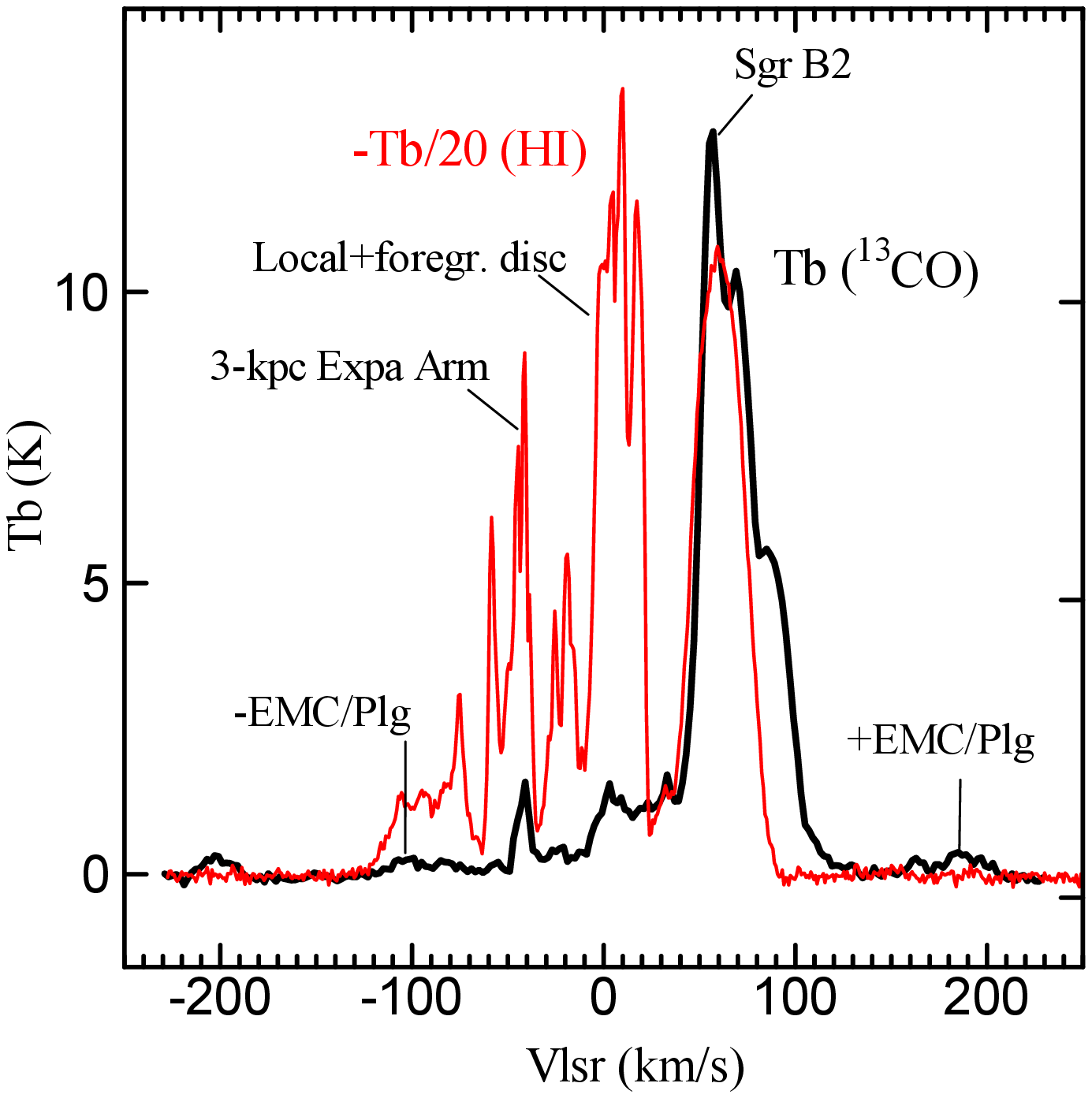}   
\end{center}
\caption{Comparison of the $^{13}$CO spectrum with the inverse HI absorption spectrum ($-\Tb/20)$ toward Sgr B2. Note the coincidence of CO and HI profiles of Sgr B2 at $\vlsr \sim 60 \ekms$.}
\label{spec13COHIB2} 
\end{figure}

\sss{Common absorption by foreground HI} 

Commonly recognized in all the panels are the absorption lines due to the local HI gas and spiral arms in the disc at velocities of $|\vlsr|< \sim 40$ \kms.
The narrow and sharp absorption at $\vlsr=-55$ \kms is due to the 3-kpc expanding ring (arm).
The negative velocity part of the EMR/EMC/Plg (marked by -EMC) is also commonly visible by absorption, which appears at different negative velocities depending on the longitudinal positions.
However, positive-velocity absorption (+EMC) by the counter (far) side of the ring/cylinder is not visible.
This means that the four radio sources are located inside the EMC.

\sss{Sgr B2} 

Sgr B2 shows a strong and wide absorption at $\vlsr \sim +60\pm 20$ \kms by Arm I and the Sgr B molecular complex. This makes a quite different appearance of the spectrum from that of Sgr B1.  

Fig. \ref{spec13COHIB2} shows comparison of the $^{13}$CO line profile with an inverse HI absorption profile toward Sgr B2.
The profiles at $\vlsr\sim +60 \ekms$ are very similar to each other, demonstrating that the HI and CO kinematics are almost identical.
The slight difference at higher velocity edge is  attributed to the difference in the observing beam sizes (CO $20''$ vs HI $145''$).

The coincidence of the velocity of recombination lines from the HII region (+64 \kms) \cite{brown+1975} with CO and HI indicates that the HII region (Sgr B2) is embedded in Arm I and Sgr B2 molecular complex.
This is consistent with the mutual close locations of HI and HII gases in Sgr B2 as derived from high-resolution absorption study with the VLA \cite{lang+2010}.

The HI spectrum shows a clear-cut, shoulder-like absorption feature at $\vlsr \sim -110 \ekms$ due to HI in the approaching half of the EMR/EMC/Plg, while no positive-velocity absorption by the receding side of EMR is seen.
On the other hand, the CO line is seen both in the negative and positive sides in emission.
From these we may conclude that Sgr B2 and associated molecular cloud is located inside the EMC/EMR/Plg.

\sss{Sgr B1} 

Spectrum of Sgr B1 is similar to that of Sgr B2 at $\vlsr\le \sim 30 \ekms$.
Arm I's absorption is seen at $\vlsr\sim 40 \ekms$, indicating the association with the Arm.
However, the intensity is much weaker than that for B2 because of the lower continuum brightness. 

\sss{Sgr A} 

The absorption spectrum toward Sgr A is as deep as several thousand degrees, and might be significantly contaminated by continuum emission of the surrounding strong radio features through the side lobes.
Nevertheless, the spectrum is very similar to that observed in Sgr A East by the VLA \cite{lang+2010}.
Considering the contamination carefully, we may recognize some prominent absorption features.

The wide absorption profile at $\vlsr \sim +50 \ekms$ is due to the "50-\kms cloud" near Sgr A East \cite{tsuboi+1999,tsuboi+2009}, which may partially represent Arm III rotating rapidly around Sgr A \cite{sofue1995a}.

The absorption features of Arms I and II toward Sgr A, supposed to appear at $\vlsr\sim -30$ and $+50$ \kms, respectively, are contaminated by the disc absorption and 50-\kms cloud, and cannot be used to discuss the Arms' location.

\sss{Sgr C}

Sgr C shows HI absorption at $\vlsr \sim -60$ \kms, coincident with its recombination-line velocity \cite{pauls+1975,anantha+1989}.
This indicates that the source is an HII region embedded in Arm II at the same radial velocity.
Unfortunately, it is close to the 3-kpc arm at $-50$ \kms at the longitude.
The spectrum also shows sharp absorption at $\sim -115$ \kms due to Arm I, indicating that Sgr C is in the far side of Arm I.
This suggests that Arm I and II compose a symmetric pair about Sgr A.
The wide and deep absorption at $\sim -150 \ekms$ is due to the approaching side of the EMR, while no positive-velocity EMR absorption is found.
This means that Sgr C and Arm II are inside the EMR/EMC.

\sss{Sgr D} 

The three SNRs show wide HI absorption on the LV diagrams over all the velocity range, indicating that they are beyond or inside the CMZ near the tangential velocity area.
Fig. \ref{lv-absorption} shows enlarged LV diagrams across the three SNRs.
G1.1-0.1 is visible as a weak and broad absorption from $\vlsr\sim -120$ to $+210$ \kms, which is overlapped by a sharper absorption due to the foreground HII region at $\vlsr\sim -20 \ekms$. 
G1.0-0.2 is a pure SNR, showing a wide absorption over all the velocity range.
SNR absorption toward G0.9+0.1 is not so clear, because the overlapped HII region shows deep absorption over the whole velocity range.
This indicates that HII region at G0.9+0.1 is located near the tangential region of CMZ or beyond the CMZ.

\sss{Sgr E}

Sgr E (E1 and E2) is not bright enough to produce HI absorption.  
We can recognize only a weak absorption-like region toward E1 at $l\sim -1\deg.05$ on the HI Arm I in the LVD (Fig. \ref{lv-absorption}) coinciding with a CO bright clump with $\vlsr\sim -180 \ekms$. 
This is consistent with the idea that Sgr E1 is a tangential direction of the turning point on the sky of SF ring and molecular Arm I to II.

\subsection{Symmetric locations of Arm I and II}
 
The 3D motion of Sgr B2 from the parallax and radial velocity indicates that Sgr B2 is in the near side of Sgr A. 
Sgr B2 is associated with the molecular complex at $(l,\vlsr) = (0\deg.65,+62 \ekms)$ and is absorbed by HI at the same $(l,v)$ position.
This velocity coincides with that of Arm I at this position, which indicates that Arm I is also located at the same position as the Sgr B molecular complex and Sgr B2 it self.
This means that Arm I is in the near side of Sgr A.   
On the other hand, Sgr B2 is not absorbed by Arm II at $\vlsr \simeq +100 \ekms$ as shown by Fig. \ref{HIabsspec}.
This means that Arm II and Sgr C is in the far side of Arm I and Sgr B2.
 
However, the location of Arm II with respect to Sgr A must be determined in another way.  
For this, we refer to the result of the method making use of the ratio of the CO-line emission to OH absorption of the radio continuum disc, which has indicated that Arm II is in the far side of Sgr A \cite{sawada+2004,sofue2017}.
So, we here conclude that Arm II with Sgr C is located in the far side of Sgr A, opposite to Arm I and Sgr B2.
Namely, Arm I and II compose a symmetric pair of arms around Sgr A, making part of the star forming radio belt connecting Sgr B1, B2, C and SNRs in the direction of Sgr D.

\subsection{Face-on view of the GC}

\begin{figure*} 
\begin{center}    
\includegraphics[width=10cm]{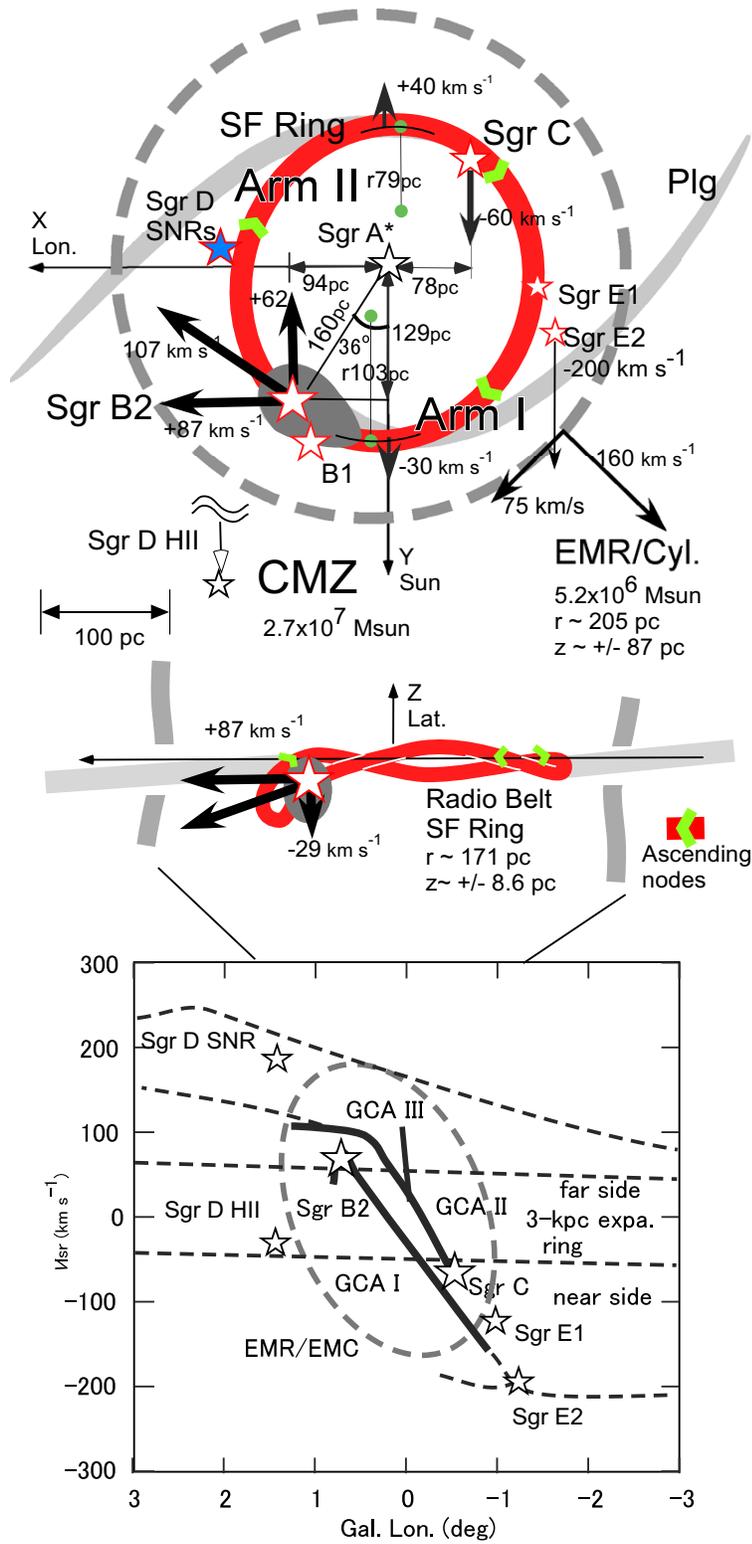}
\end{center}
\caption{Illustration of relative locations of Sgr A, B, C and GC Arms I and II . [Top] Face-on view, [middle] edge-on view (on the sky), [bottom] LV diagram.}
\label{illust}  
\end{figure*}

Fig. \ref{illust} illustrates the obtained face-on and edge-on orientations of the Arms and radio sources, which also satisfy the condition to produce the rigid-body ridges in the LV diagram.
The derived parameters are listed in table 2.
In this view, Arms I and II comprise a global ring structure.
The slight offsets of $l$-axis intersections of LV ridges of Arms I and II of $\pm \sim -20$ and $\sim +30 $ \kms, respectively, allow for a slightly oval orbits.

The arms are tilted and bent about the Galactic plane, and the top of Arm I with Sgr B2 is moving downward of the disc.
This motion is consistent with the bent structures of the Arms and the radio belt shown in Fig. \ref{belt}.

\begin{table*}  
\caption{Kinematic and structural parameters$^\dagger$ of the CMZ, CHZ,  Arms I and II, EMR/EMC/{Plg}, Sgr A$^*$, B, and C} 
\begin{tabular}{lll}
\hline\hline
Structure & Parameter&Remarks  \\ 
\hline
CMZ \dotfill 
    & $h_r\sim \pm 207$ pc 
    & Plateaued prof., $l=-1\deg.1$ to $+1\deg.8$ ($-158$ {\rm to} +257 pc)\\
    & $h_z\sim \pm 28$ pc & Gaussian-like profile\\
    & $M_{\rm C}$(H$_2$)$\sim 2.3\times 10^7 \Msun$ (incl. metals by $\mu=1.4$)&  $\Xco(0)=0.51\times 10^{20}$ \xcounit\\
    & $M_{\rm C}$(HI inside CMZ)$\sim 2.3\times 10^6\Msun)$ (ibid)\\
    & $\fmol ({\rm total\ mass})\sim 0.91$
    & For total corrected masses inside CMZ\\
    & $\fmol ({\rm column}) \simeq 0.9-0.95$ 
    & For column densities in the Galctic plane \\ 
    \hline
CHZ  \dotfill
    & $h_r \sim \pm 322$ pc
    & Nearly exponential: $l=-2\deg.0$ to $+2\deg.5$\\
    & $h_z\sim \pm 71$ pc
    & Exponential profile\\
    & $M_{\rm C}$(HI)$\sim 6.9
    \times10^6 \Msun$ (ibid) &Disc eliminated + fan-correction\\
    & CHZ/CMZ mass ratio $\sim 0.25$& Note the larger CHZ area than CMZ\\ 
    \hline
EMR/EMC/{Plg}  \dotfill

    & $r\sim 205$ pc
    &\cite{sofue2017}\\
    & $h_z\sim \pm 87$ pc
    &\\
    & $v_{\rm expa.}\sim 160 \ekms$ \\
    & $v_{\rm rot}\sim 75 \ekms$ \\
    & $M\sim 4.5\times 10^6 \Msun $ &$\Xco(0)$ as above \\
    & EMC to CMZ mass ratio $\sim 0.2$ \\
    
    \hline
                      
Sgr A$^*$  \dotfill
    &G$-0.05-0.05$ \\
    & $(X, Y, Z)=(0, 0, 0)$ & assumed\\
    & $(\vlsr, V_l, V_b)=(0,0,0)$ & assumed\\
    \hline
    
Sgr B2  \dotfill
    & G$+0.66-0.04$  \\
    & $ X=94$ pc 
    & positive toward increasing long.\\
    & $ Y\simeq 129$ pc 
    & positive toward Sun \\
    & $ Z=-5.7$ pc 
    &\\
    & $R\simeq 160$ pc 
    &\\
    & $\theta \simeq 36\deg$ &from $Y$ axis \\
    & $\vlsr=+62.3 \ekms $ \\
    & $V_l \simeq +87.4$ \kms &Proper motion \\
    & $V_b \sim -29$\kms & Proper motion \\
    & $V_{\rm rot}\simeq 107$ \kms &$[\vlsr ^2 + V_l ^2]^{1/2}$ \\
    & $V_{\rm orb.}\simeq 111$ \kms &$[\vlsr ^2 + V_l ^2 + V_b ^2]^{1/2}$ \\
    &$i=-14\deg.2$ & orbit inclination\\
    &$PA=108\deg$ & direction on the sky\\
    \hline

Sgr C  \dotfill
    & G$-0.56-0.08$ \\
    & $X=-80$ pc\\
    & $Z=-11$ pc &\\
    & $\vlsr=-60 \ekms$ \\
    \hline
    
Arm I \dotfill
     &$dv/dl\sim 147 $ \kms deg$^{-1}$ \\
    & $r_{\rm curv.}\sim 103$ pc 
    & \\
    & ${\vlsr}({l=0\deg})\sim -30 \ekms $\\
    & $l({\vlsr=0 \ekms})\sim 0\deg.1$\\
    \hline

Arm II \dotfill
    & $dv/dl\sim 193 $ \kms deg$^{-1}$\\
    & $r_{\rm curv.}\sim 79$ pc 
    &\\
    & ${\vlsr}({l=0\deg})\sim +40 \ekms$ \\
    & $l({\vlsr=0 \ekms})\sim -0\deg.15$\\
\hline
Warping ring \dotfill
    & Radius $1\deg.2\sim 171$ pc 
    & Edge-on view draws double infinity, $\dinfty$.\\
    & Amplitude $\delta z\sim \pm 0\deg.06 \sim \pm 8.6$ pc\\ 
    & Ascending nodes at $l\sim -0\deg.7, \ +0\deg.9, \ -0\deg.8$\\
\hline
\end{tabular}\\
$\dagger$ \red{GC distance is assumed to be $R_0=8.18$ kpc \cite{gravity+2019}.}
\label{tab2}
\end{table*}

We stress that the here obtained 3D orientation is a unique solution, when we consider the following observational facts.\\
(i) Rigid-body Arms I and II on the LVD, indicating nearly circular orbits.\\
(ii) HI absorption toward Sgr A, B and C. \\
(iii) 3D motion of Sgr B2 with respect to Sgr A$^*$ inferred from the proper motions and radial velocities.\\
The line-of-sight location and 3D motion of Sgr B2 with respect to Sgr A have been a long-standing issue.  
The present conclusion is consistent with most of the current works \cite{sofue1995a,tsuboi+1999,sawada+2004,ryu+2009,henshaw+2016,sofue2017,henshaw+2022}.

However, some authors suggest far-side location of Sgr B2 \cite{rohlfs+1982,molinari+2011,okatake+2022}. 
One could explain it by ignoring condition (i) by elongating the ring nesting Sgr B, C, Arms I and II in such a way that the major axis runs in the direction from top-left to bottom right in Fig. \ref{illust}. Sgr B must be located on the ellipse in the far side of $X$ (Gal. Lon.) axis, and Sgr C in the opposite position. In this case, we must abandon the assumption that the 3D velocity vector of Sgr B represents a nearly circular motion around Sgr A, but it is orbiting along a highly elongated elliptical orbit, as suggested by numerical simulation of gas flow in a bar potential \cite{tress+2020}. A problem about this orbit is that the gas is oscillating almost radially, so that the LVD draws an ellipse with maximum velocities near the $Y$ axis ($l\sim 0\deg$), contradicting the observed straight rigid-body ridge.  

\section{Summary and discussion} 

We have obtained detailed comparison of the CO and HI maps and LV diagrams, and showed that CO features are closely associated with HI.
We further examined trigonometric positions and kinematics of the radio continuum sources Sgr A$^*$ and B2.
We thus derived a 3D structure of the GC, and determined the parameters as presented in Fig. \ref{illust} and table \ref{tab2}. 

We summarize the results as follows:
\begin{itemize}

\item The CMZ is defined as a region with a plateaued distribution of the CO line intensity in the longitudinal direction, which encloses an area from $l=-1\deg.1$ ($-157
$ pc) to $+1\deg.8$ (257
pc).
The plateaued profile indicates a ring-like gas distribution.
\item CMZ has a Gaussian-like vertical profile with the $e$-folding $z$ half-thickness of $\pm 0\deg.2$ (29 pc).
\item The CHZ (Central HI Zone) composes an HI disc extending from $l=-2\deg$ to $+2\deg.5$ ($-290$ to $+357$ pc), showing exponential concentration toward the GC.
\item  CHZ has an exponential vertical profile with scale height $\sim \pm 0\deg.5 (\pm 71)$ pc, much thicker than the CMZ. 
\item Total HI mass inside CMZ is abput 10\% of the molecular gas, and the molecular fraction by column density is as high as $f_{\rm mol}\simeq 0.9-95$. 
\item The GC molecular Arms I and II are traced by a radio continuum belt, which draws a double-infinity $\dinfty$ shape on the sky. The belt is understood as an edge-on projection of a warping SF ring having three ascending nodes.
\item Tight correlation is found between the HI and CO structures in the LVD. Molecular Arms I and II show up clearly in HI in the LVDs.  
\item HI absorption toward continuum sources Sgr A, B1, B2 and C was used to determine the line-of-sight locations of the Arms with respect to the radio sources. 
\item The $dv/dl$ method was used to determine the curvatures of Arms I and II.   
\item The 3D motion of Sgr B2 relative Sgr A$^*$ inferred from the proper motion and radial velocity restricts the orbit direction and position of B2 about A$^*$. 
\item Arms I and II are the major composition of CMZ close to the galactic plane within $\sim \pm 20$ pc. 
Each arm is much narrower, as thin as $\sim 4-10$ pc. 
\item Slight offsets of the LV ridges from a pure rigid body allow for mildly oval or elliptical orbits of the Arms. 
\item EMR/EMC/Plg is present in HI, most clearly visible at negative high latitudes at $b\sim -0\deg.5$ as a round LV ellipse.
\item EMR/EMC/Plg shows quite different properties from CMZ.
It has only an order of magnitude lower mass ($\sim 4.5\times 10^6\Msun$) than Arms I and II, and vertically extends over $\sim 174(\pm 85)$ pc, 
indicating that the EMC cannot be the source to feed gas to CMZ. 
\end{itemize}

The derived parameters will be useful for putting constraint on theoretical models of the Galactic Center, particularly on 3D models.
The here presented view may be modified, so that the major parts of the Arms are interacting with the outer arm structures in such a sense that the HI arms are winding in from outside and are linked smoothly to the ring. Such dual flows, comprising a central ring and two spiral fins or dark lanes, are often observed in spiral galaxies \cite{combes+2004}, and are simulated in numerical computations \cite{fux1999,kim+2011,salak+2017,shin+2017,ridley+2017,sormani+2018,sormani+2019,tress+2020,armi+2020,hatch+2021}. 

It may be also worth to consider the explosion model for EMC (e.g., Sofue 2017) more seriously based on the 3D parameters.
The properties such as an order of magnitude less mass than CMZ, several times larger thickness than Arms I and II, and the round LVD (not parallelogram) at high latitudes, can be explained by the explosion model. 
The explosion hypothesis is also consistent with a variety of energetic shocks and outflows observed in radio, X and gamma rays \cite{kataoka+2018,ponti+2019,heywood+2019}, which are a natural consequence of the feedback \cite{sofue2020} of accretion of gas to the nucleus by the bar models. \\

\vskip 5mm
\noindent{\bf ACKNOWLEDGEMENTS}:
\vskip 2mm \noindent The author expresses his grateful thanks to Prof. Tomoharu Oka and his collaborators for the CO-line data taken with the Nobeyama 45-m Telescope, to Prof. Naomi McGraugh-Griffith and the collaborators for the HI line data taken with the ATCA, and to Prof. I. Heywood and the collaborators for the MeerKAT 1.28 GHz image of the GC.
The data analysis was performed at the Astronomical Data Center of the National Astronomical Observatory of Japan. 
 
\vskip 3mm
\noindent{\bf DATA AVAILABILITY}:
\vskip 2mm\noindent The CO-, HI-line and radio continuum data were downloaded from the url,\\
https://www.nro.nao.ac.jp/$\sim$nro45mrt/html/results/
data.html;\\
https://www.atnf.csiro.au/research/HI/sgps/
GalacticCenter/Home.html, and\\
https://archive-gw-1.kat.ac.za/public/repository/10.48479/
fyst-hj47/index.html.

\vskip 3mm
\noindent{\bf NO CONFLICT OF INTEREST}:
\vskip 2mm \noindent The author declares that there is no conflict of interest.

\def\apj{ApJ} \def\apjs{ApJ. Suppl.} \def\apjl{ApJ. L.} \def\aap{A\&Ap} \def\aj{AJ} \def\mnras{MNRAS} \def\pasj{PASJ} \def\araa{ARA\&Ap}

\newpage
\begin{appendix}
\section{Galactic-disc elimination by low-Velocity cutting (DEV)}\label{Appendix}  
 
\begin{figure} 
\begin{center}      
\includegraphics[width=7.5cm]{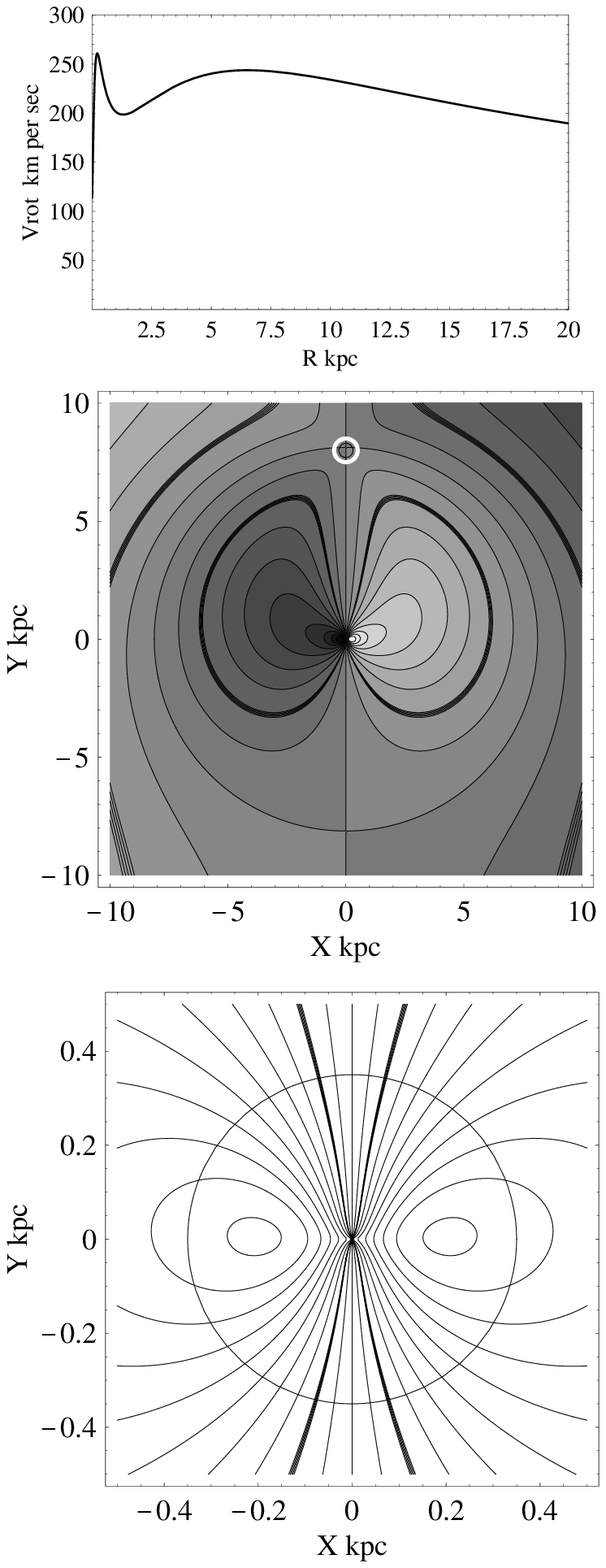} 
\end{center}
\caption{{
[Top] Model rotation curve of the Milky Way mimicking the observed curve (Sofue 2013).
[Middle] Radial velocity field in the Galactic plane, and [bottom] Same, but enlarged near GC.
The Sun is at $(X,Y)=(0,8)$ kpc as marked by a circle. 
Contours are drawn every 25 \kms. 
By DEV, the emission from the fan-shaped region near $Y$ axis with $|\vlsr|< 50$ \kms is eliminated. 
This also eliminates a fraction of $2\phi/\pi$ of the emission of the central disc, where $\phi=\sin^{-1}(V_{\rm cut}/\Vrot)$. 
So, the final mass must be corrected by a factor of $1/(1-2\phi/\pi)$.}}
\label{vfield}  
\end{figure}

Contamination of the CO and HI line emissions from the fore- and background Galactic disc causes crucial over-estimation of the integrated intensity of the GC objects due to radial-velocity degeneracy.
It comes from the narrow fan-shaped region near $Y$ axis  between the thick contours at $\vlsr=\pm \vcut=50 \ekms$ in Fig. \ref{vfield}, for example.
This component can be eliminated using the DEV method by cutting the emission at $\vlsr|\le \vcut$, yielding Map C.
Map C is used to calculate the mass $M_{\rm C}^0$ of the object.

However, the procedure also eliminates a fan-shaped area of the central disc itself (CMZ, CHZ) at $\sin \phi \le v_{\rm cut}/\Vrot$, where $\phi$ is the azimuth angle from $Y$ axis. 
For rotation velocity of $\Vrot \sim 150 \ekms$ \cite{sofue2013} the half-fan angle is $\phi\sim 19.5\deg$. 
Namely, a fraction of $2\phi/\pi\sim 0.22$ of the supposed total mass of CMZ is missing in the DEV map.
This over-eliminated flux can be recovered by multiplying a correction factor, $\gamma=1/(1-2\phi/\pi)$, to the calculated mass $M_{\rm C}^0$ from Map C by Eq. (\ref{mass3}), yielding the final corrected mass, $M_{\rm C}$,
on the assumption that the disc is axisymmetric. 
 
\begin{figure} 
\begin{center}      
\includegraphics[width=8cm]{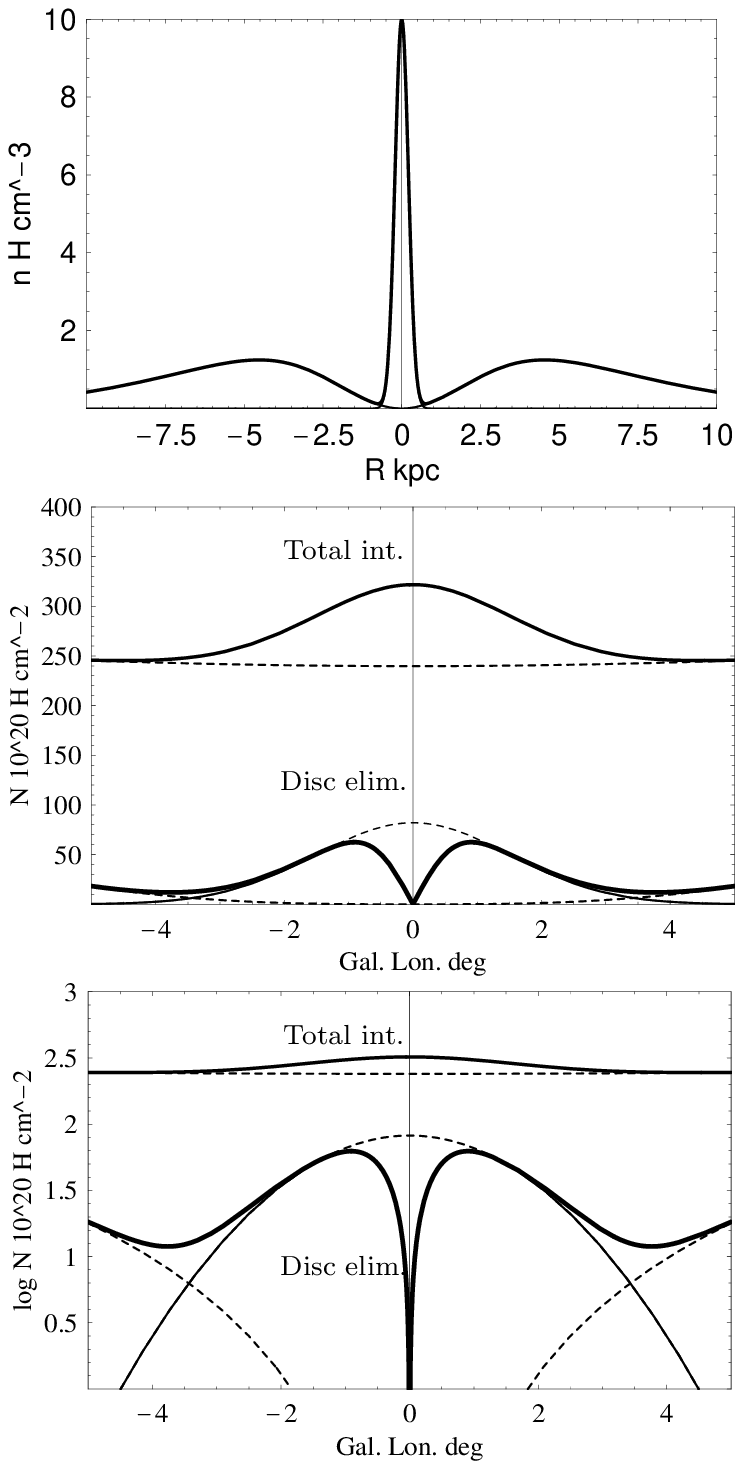}   
\end{center}
\caption{{
[Top] Model HI density distribution.
[Middle] Calculated HI column density in linear scaling before (upper line) and after DEV (lower line). Here the rotation velocity is assumed to be constant at 150 \kms in the gC.
[bottom] Same, but in semi-logatithmic (bottom) scaling.  
The lower plot of the bottom panel may be compared with the observed HI profile in Fig. \ref{profile}}}
\label{model}  
\end{figure}

We here simulate an expected profile of the HI column density before and after the DEV.
We assume a rotation curve and gas density distribution as shown in Fig. \ref{vfield}.
The central disc representing CHZ and an extended ring-like disc is shown in Fig. \ref{model}, which is expressed by 
\begin{equation}
n_{\rm CHZ}=10\ \exp(-(R/r_{\rm CHZ})^2)\ {\rm H\ cm^{-3}},
\end{equation}
and
\begin{equation}
n_{\rm disc}=2\ [\exp(-(R/r_{\rm d1})^2) - \exp(-(R/r_{\rm d2})^2)]\ {\rm H\ cm^{-3}}.
\end{equation}  
Here, 
$r_{\rm CHZ}=300$ pc is the scale radius of CHZ, and $r_{\rm d1}=8 $ kpc is the radius of the outer ring-like HI disc, and $r_{\rm d2}=3$ kpc denotes the radius of HI-deficient region in the inner Galaxy. 

Calculated longitudinal profiles of the column density are shown in Fig. \ref{model}.  
The upper profiles in each panel indicates total emission of the galactic disc and CHZ integrated over the full velocity range.
The lower profiles show the disc-eliminated result, where the low-velocity cut at $|\vlsr|=50 \ekms$ has been applied. 

In the logarithmic presentation (bottom panel), the full-velocity range integration yields an almost flat profile, so that CHZ is buried in the disc's profile and is hard to be detected.
On the other hand, the profile after low-velocity cut effectively suppresses the extended disc, and CHZ shows up clearly.
The lower profile in the bottom panel may be compared with the observed HI profile in Fig. \ref{profile}.

Note, however, that the GC disc itself also suffers from the cut at  $|l| \le \sim1\deg$ (in case of $\Vrot=150-200$ \kms), exhibiting a V-shaped sharp drop.  So, the intensity maps and profiles must be regarded as to represent minimal structures in the GC.
This over-eliminated flux can be recovered by the correction facto as described above.

\end{appendix} 
\end{document}